\pdfoutput=1 
\documentclass[pre,twocolumn,showpacs,floatfix,nofootinbib,groupaddress,showkeys]{revtex4-1}

\usepackage{amsmath}
\usepackage{amsfonts}
\usepackage{amssymb}
\usepackage[margin=0.75in]{geometry}
\usepackage{graphicx} 
\usepackage{cleveref}
\usepackage{bm}
\usepackage{color}
\usepackage{ifpdf}
\ifpdf
\fi
\usepackage{feynmp}
\usepackage[outdir=./]{epstopdf}
\newcommand{\ket}[1]{|#1\rangle}

\renewcommand{\[}{\begin{equation}}
\renewcommand{\]}{\end{equation}}
\renewcommand{\Re}{\operatorname{Re}}
\renewcommand{\Im}{\operatorname{Im}}

\begin{document}

\begin{fmffile}{microe}

\bibliographystyle{apsrev}
\title{Microemulsions in the driven Widom-Rowlinson lattice gas}
\author{Maxim O. Lavrentovich}
\email{lavrentm@gmail.com}
\affiliation{Department of Physics \& Astronomy, University of Tennessee, Knoxville, Tennessee 37996, USA}
\author{Ronald Dickman}
\email{dickman@fisica.ufmg.br }
\affiliation{Deparamento de F\'isica and National Institute of Science and Technology for Complex Systems, ICEx, Universidade Federal de Minas Gerais, C. P. 702, 30123-970 Belo Horizonte, Minas Gerais, Brazil}
\author{R. K. P. Zia}
\email{rkpzia@vt.edu}
\affiliation{Center for Soft Matter and Biological Physics, Department of Physics, Virginia Polytechnic Institute \& State University, Blacksburg, Virginia 24061, USA}
\affiliation{Physics Department, University of Houston, Houston, Texas 77204, USA}
\begin{abstract}
 An investigation of the two-dimensional Widom-Rowlinson lattice gas under an applied drive uncovered a remarkable non-equilibrium steady state in which  uniform stripes (reminiscent of an equilibrium lamellar phase) form perpendicular to the drive direction [R. Dickman and R. K. P. Zia, \textit{Phys. Rev. E} \textbf{97}, 062126 (2018)].  Here we study this model at low particle densities in two  and three dimensions, where we find a disordered phase with a characteristic length scale (a ``microemulsion'') along the drive direction. We develop a continuum theory of this disordered phase to derive a coarse-grained field-theoretic action for the non-equilibrium dynamics. The action has the form of two coupled driven diffusive systems with \textit{different} characteristic velocities, generated by an interplay between the particle repulsion and the drive. We then show how fluctuation corrections in the field theory may generate the characteristic features of the microemulsion phase, including a    peak in the static structure factor corresponding to the characteristic length scale.  This work lays the foundation for understanding the stripe phenomenon more generally.         \end{abstract}
\maketitle

\section{Introduction and Background}

\begin{figure}[htp]
\centering
\includegraphics[width=0.45\textwidth]{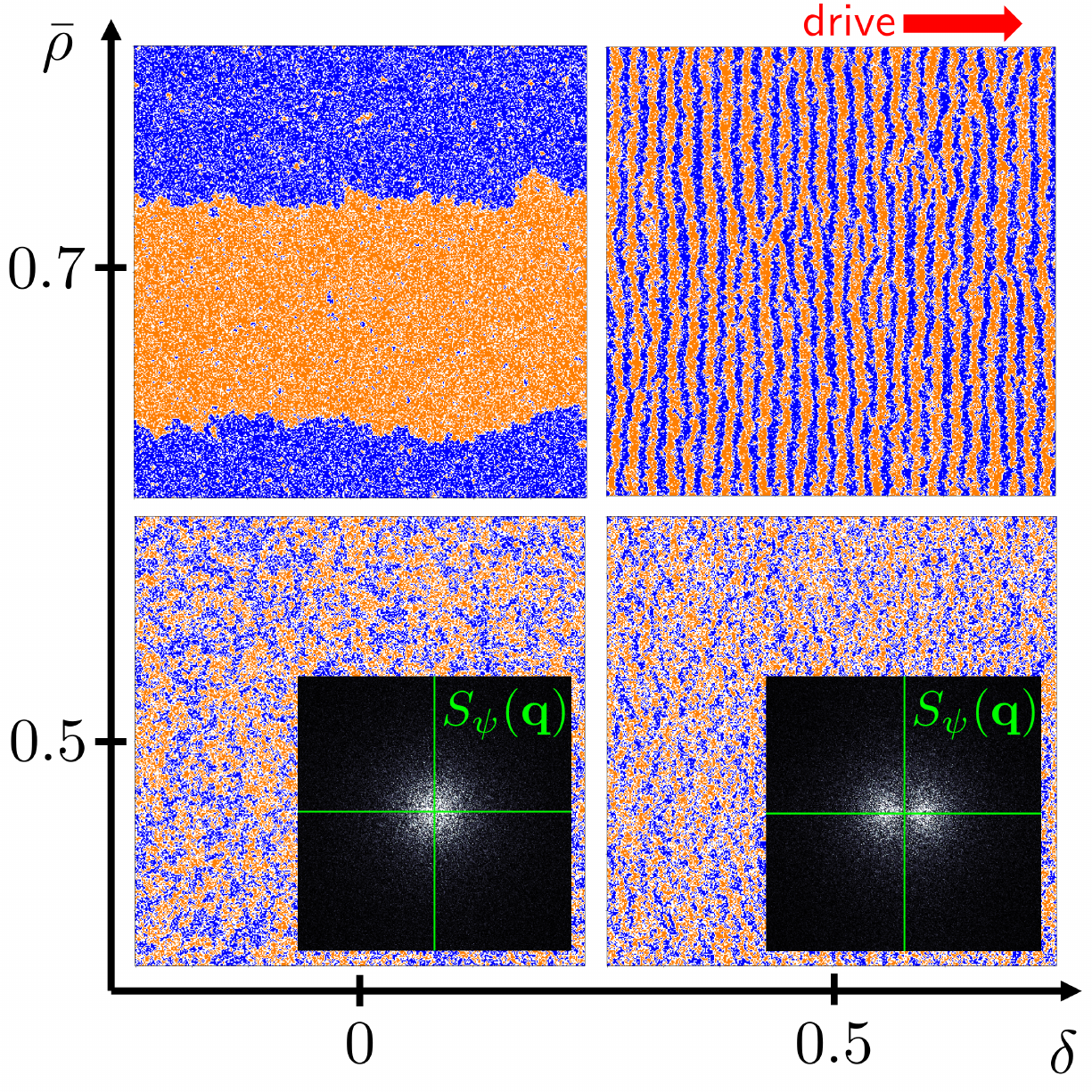}
\caption{\label{fig:intro} Snapshots of steady states of the driven Widom-Rowlinson lattice gas (DWRLG) for a $L \times L$ system with $L=400$ with equal mixtures of blue and orange particles. At high densities ($\bar{\rho}=0.7$), the  mixture phase separates either completely without a drive ($\delta=0$) or into uniform stripes with a drive ($\delta=0.5$).  At lower densities ($\bar{\rho}=0.5$), the particles remain mixed. In the presence of a drive, in the direction given by the red arrow, the disordered phase has a characteristic length scale as can be observed from the static structure factors  $S_{\psi}(\mathbf{q})$ (the insets). Note that the $\delta=0$ case has a single peak at the origin, while the $\delta=0.5$ case has two maxima away from the origin along the drive direction.} 
\end{figure}

Modulated or spatially-patterned phases abound in and out of equilibrium: Block copolymers, magnetic thin films, type-I superconductor films 
in applied magnetic fields, cholesteric liquid crystals  \cite{brazovskiiLC}, and lipid mixtures may all exhibit thermodynamically stable, 
patterned equilibrium phases \cite{ModPhasesOverview}. On the other hand, driven granular systems \cite{granular1}, Rayleigh-B\'enard convection rolls  
\cite{SwiftHohenbergConvection}, and other out-of-equilibrium systems exhibit spatially-modulated, dynamical steady states \cite{CrossHohenberg}, 
as well. In  these systems, when an equilibrium, free-energy-based treatment is appropriate, one may often define a coarse-grained scalar order 
parameter $\psi(\mathbf{r})$ which, in the ordered phase, has some spatial modulation with a characteristic wavenumber $q^*=2\pi/\lambda^*$ 
describing the pattern size $\lambda^*$. A common ordered phase consists of stripes (or slabs, in 3D) with $\psi(\mathbf{r}) \propto \cos (q^*x)$ 
for describing a structure periodic in some $\hat{\mathbf{x}}$ direction. 

In the disordered phase, with $\langle \psi \rangle=0$ on average, the characteristic pattern size may also show up as a peak in the 
static structure factor at $q^*$. Such a peak is observed in scattering intensity distributions of oil-and-water mixtures, which may be 
treated with a phenomenological free energy of the kind considered here \cite{originalmicro}. We therefore refer to the   ``structured''  
disordered phase as a microemulsion. 
Such a phase is characterized by clustering at a particular length scale $\lambda^*$, but without any long-range, ordered patterning.  
Such phases may contain droplets or have a bicontinuous, disordered structure.  The presence of a characteristic $q^*>0$ in the disordered 
phase also strongly modifies the phase behavior, as the thermal fluctuations of the order parameter $\psi$  occur predominantly 
in a non-zero momentum ``shell'' $|\mathbf{q}|=q^*$.    

For the systems admitting an equilibrium treatment, the origin of the special scale $q^*$ may come from some competing interactions, 
such as a coupling of a lipid membrane   composition to the membrane curvature \cite{andelmanmembrane} or be set by particular 
boundary conditions \cite{SwiftHohenbergConvection}. Once the scale $q^*$ has been identified, such systems  near the pattern-formation 
transition are described by a phenomenological, coarse-grained free energy, first analyzed by Brazovskii and coworkers \cite{brazovskii,muratov}.    
In this work, we explore a system where the scale $q^*$ develops unexpectedly and no coarse-grained free energy is \textit{a priori} available 
due to the explicitly out-of-equilibrium state of the system: The patterns develop in a  phase-separating binary mixture of particles with 
purely repulsive interactions under an applied drive \cite{DRWLG1}.  We will show that the origin of $q^*$ is  subtle in this case, and 
results from an interplay between the repulsive interactions and the applied drive.

Strongly-driven physical systems exhibit a range of out-of-equilibrium, pattern-forming phenomena.  For example, laning or striped behavior is 
known to occur in a wide range of systems, including vibrated granular mixtures with varying friction coefficients \cite{granvib1}, 
driven polymer blends \cite{polblend}, and  binary plasmas \cite{plasmas}. In most of these systems, the stripes form parallel to the drive direction.  Moreover, the size of the stripes is typically set by either the microscopic constituents of the model, or by the system size \cite{Mitra2018}. Otherwise, non-trivial spatial modulations appear to emerge in systems with long-range particle-hole exchanges \cite{Mohanty2016}. By contrast, with only local dynamics, our model exhibits ordering at scales between the lattice spacing $\ell$ and the system size $L$, admitting a coarse-grained, hydrodynamic description.

Here we study a  stripe formation phenomenon in a lattice gas model under a steady 
applied drive. We find non-equilibrium steady states with similar characteristics as the equilibrium  phases described by the Landau-Brazovskii 
free energy for modulated phases. In particular, there is a phase transition between an ordered modulated (striped) phase at high particle 
density $\bar{\rho}$ and a disordered ``microemulsion'' phase at low densities in which the static structure factor  has a characteristic peak. 
We  develop a coarse-grained field theory for this process and show how our model captures two important limiting cases: The phase-separating 
binary mixture and a driven diffusive lattice gas. Similarly, many model systems also exhibit striped patterns when driven far from equilibrium. 
Examples include the venerable Ising system \cite{I25} and the 3-state Blume-Emery-Griffiths model \cite{BEG71,P52}, especially when cast in the lattice 
gas language \cite{Fowler36,Lacher37,LY52,Temp54,BR71,LS75}. Subjected to a direct or a random drive \cite{KLS1,KLS2,LSZ89,DRWLG1,GLMS90,CGLV91,SZ91}, as well as various boundary conditions, 
these systems display stripes oriented parallel or perpendicular to the drive \cite{VLZ89,BSZ91,SHZ92,BSZ93,LZ97}. 

\begin{figure}[htp]
\centering
\includegraphics[width=0.45\textwidth]{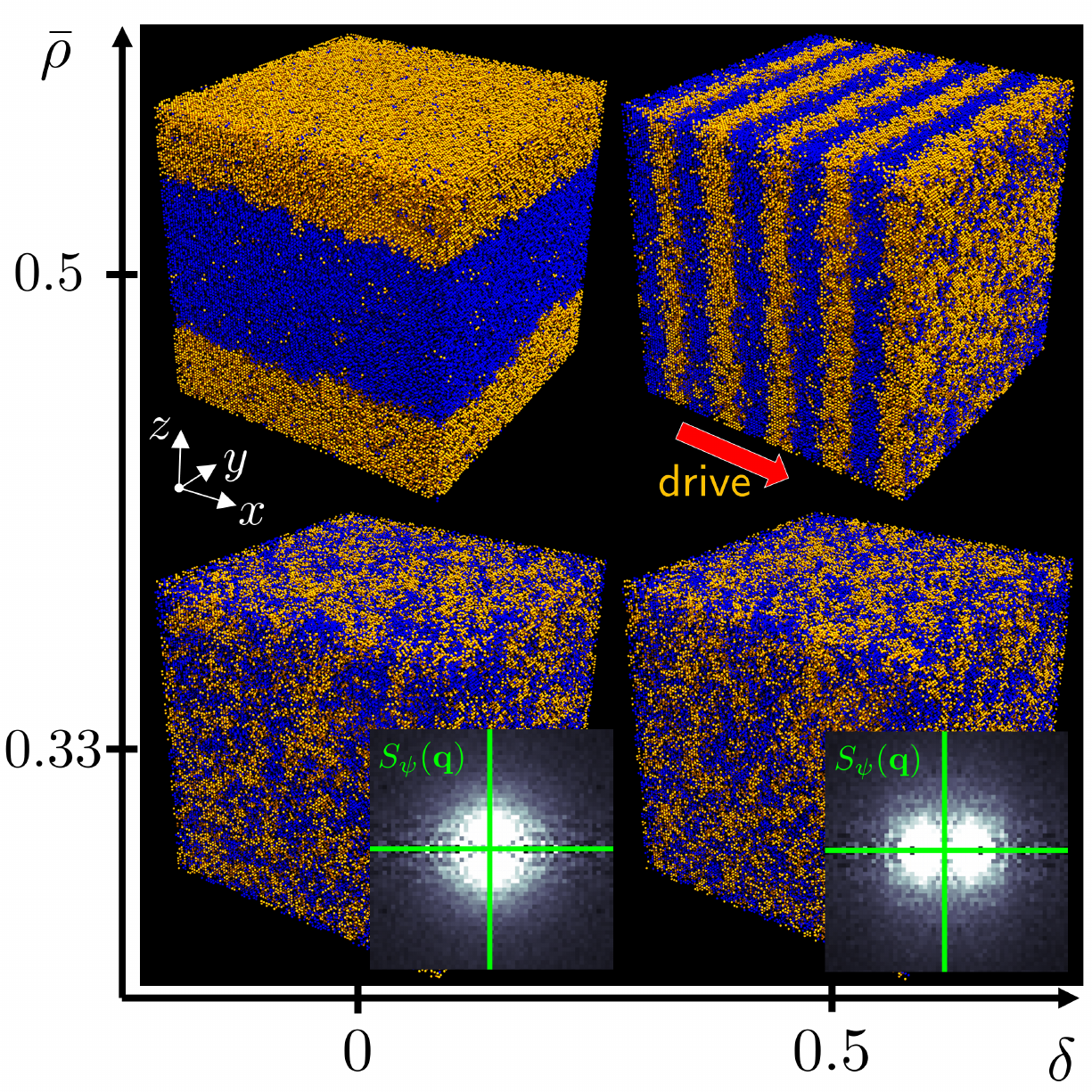}
\caption{\label{fig:intro3d} Snapshots of steady states of the DWRLG for a $L \times L \times L$ system with $L=100$ with equal mixtures of blue and orange particles. At high densities ($\bar{\rho}=0.5$), the binary mixture phase separates either completely without a drive ($\delta=0$) or into  slabs with a drive ($\delta=0.5$). The drive direction, $\hat{\mathbf{x}}$, is shown with a red arrow.  At lower densities ($\bar{\rho}=0.33$), the particles remain mixed. The static structure factors $S_{\psi}(\mathbf{q})$ for the disordered phases are shown as a function of $\mathbf{q}=(q_{\parallel},q_{\perp})$, where we average over the two directions perpendicular to the drive.  Note   the two maxima in the drive direction  (inset on bottom right).} 
\end{figure}

Granular \cite{granstripes} and colloidal systems \cite{colloidosc1} under an oscillatory drive can also form jammed clusters with the 
stripes running perpendicular to the drive direction.   The dynamic instabilities observed in the sheared granular systems are reminiscent of the phenomenon described here \cite{granvib2,granvib22}. However, the stripes of granular particles typically coarsen over time, with the cluster sizes eventually approaching the system size. Conversely, our model shows stable structures at a characteristic scale $\lambda^*$ at the longest time scales available in our simulations ($10^8$ Monte-Carlo steps, as discussed in the next section). Moreover, fully phase-separated states are observed to breakup until the clusters reach the characteristic size.   

We consider the Widom-Rowlinson lattice gas with two species \cite{eqRWLG1}, $A$ and $B$. The particles hop on a square or a simple cubic lattice, 
subjected to excluded-volume interactions. $A$ and $B$ particles cannot occupy nearest-neighbor sites, modelling a repulsive interaction between 
species.\footnote{Whether driven or not, this system on a one-dimensional periodic lattice can be mapped (1-1) onto a ring with just one particle species
hopping from one site to a nearest neighboring vacant site, with or without bias. 
Known as  simple exclusion processes (SEPs) or  asymmetric SEPs (ASEPs), the single species systems have been extensively studied 
\cite{Spitzer70,Harris65,Liggett85,Spohn91,Liggett99}. (For a recent review, see, e.g., Ref. \cite{CMZ11}.) 
In particular, the stationary distribution is always uniform and the static properties are trivially drive-independent. 
As for the mapping, note that every allowed configuration of the Widom-Rowlinson lattice gas can be transformed into one in SEP/ASEP -- 
by deleting a single vacancy from any cluster of holes lying between an $A$ and a $B$ particle, and then relabeling all the $B$ particles as $A$'s (say).
It is straightforward to verify that, under this map, the rules for how one configuration changes to another is exactly preserved. Simulations data are entirely consistent with this picture.} The original, off-lattice version of this model with purely repulsive interactions may be mapped to a \textit{single}-component gas with 
\textit{attractive} interactions and exhibiting a vapor-liquid transition \cite{OGRW}. In the two-species case, the analog of the vapor-liquid 
transition is a phase separation of the two species when their density $\bar{\rho}$  (for equal proportions of the two species, 
$\bar{\rho}_A=\bar{\rho}_B=\bar{\rho}/2$) is larger than a critical value $\bar{\rho}^*$. At low densities $\bar{\rho}$, the particles remain 
mixed on average. These ordered and disordered equilibrium phases are shown in the left panels of Figs.~\ref{fig:intro} and \ref{fig:intro3d}  
for two- and three-dimensional systems, respectively. A key quantity that characterizes the ordering process (into periodic structures) 
is the difference between the coarse-grained 
(local) particle densities $\psi(\mathbf{r},t)=\rho_A(\mathbf{r},t)-\rho_B(\mathbf{r},t)$. We will study what happens to this density when the 
particles are ``driven'' uniformly, i.e., particle-hole exchanges in one direction are biased, as if the particles are placed in a uniform gravitational field. 
The effects of the drive are dramatic, as shown in the right panels of Figs.~\ref{fig:intro}, \ref{fig:intro3d}. At low 
$\bar{\rho}$, we find a disordered phase with a characteristic peak in the static structure factor $S_{\psi}(\mathbf{q})$ 
(associated with the ``charge field'' $\psi$) along the drive direction (a ``microemulsion'') and at high $\bar{\rho}$ we find an ordered phase of 
stripes (or slabs in 3D), reminiscent of a smectic liquid crystal. 

To understand the structure of these phases, we begin with a free energy for the equilibrium case with no drive. It has been shown that without 
the drive, the lattice gas phase separation is in the Ising universality class \cite{eqRWLG1} (with the underlying system being a diluted Ising one).
As the number of particles remains fixed, we expect the charge field $\psi$  to behave as a conserved ``magnetization,'' with the following coarse-grained 
free energy $\mathcal{F}$:
\begin{equation}
\mathcal{F}=\int \mathrm{d}\mathbf{r}\, \mathrm{d}\mathbf{r}'\psi(\mathbf{r}')G^{-1}(\mathbf{r}-\mathbf{r}')\psi(\mathbf{r})+\int \mathrm{d}\mathbf{r}\,V_I[\psi(\mathbf{r})], \label{eq:LGFE}
\end{equation}
where $G(\mathbf{r}-\mathbf{r}')$ is the Green's function   and $V_I[\psi(\mathbf{r})]$   are the higher-order interaction terms.  
For Ising systems, we expect that the relevant interaction is $V_I[\psi(\mathbf{r})]=u \psi^4$ and the (Fourier-transformed, inverse) 
Green's function is the usual $G^{-1}(\mathbf{q}) = \tau+Cq^2$, with constants $C>0$ and $\tau$.  Ignoring fluctuations, $\tau \propto \bar{\rho}^*-\bar{\rho}$ 
is our control parameter with $\tau>0$ for the disordered phase (bottom left of Figs.~\ref{fig:intro}, \ref{fig:intro3d}) and $\tau<0$ for 
the phase separated phase (top left of Figs.~\ref{fig:intro}, \ref{fig:intro3d}). The particle conservation law would show up in the 
equation of motion for $\psi$, which must be of the conserved   (model B \cite{HohenbergHalperin}) form given by
\begin{equation}
\partial_t \psi(\mathbf{r},t)= D \nabla^2 \, \frac{\delta \mathcal{F}}{\delta \psi}+\xi(\mathbf{r},t), \label{eq:LEintro}
\end{equation}
with $\xi(\mathbf{r},t)$ a Gaussian, conserved noise with correlations satisfying the fluctuation-dissipation theorem: 
$\langle \xi(\mathbf{r},t) \xi(\mathbf{r}',t) \rangle = -2 k_BTD\nabla^2\delta(\mathbf{r}-\mathbf{r}')\delta(t-t')$, with $D$ 
a diffusion constant, $k_B$ the Boltzmann constant, and $T$ the temperature. The mobility $D$ and temperature $T$ depend on 
the specific microscopic lattice rules, as well as the coarse-graining procedure. The important point to make here is that, 
without a drive, there is nothing to set the preferred scale $q^*$. The disordered phase $(\tau>0)$ has a structure 
factor (i.e., the Fourier-transform of the equal-time two point correlation) $S_{\psi}(\mathbf{q}) \propto (\tau+C q^2)^{-1}$ 
peaked at the origin $q=|\mathbf{q}|=0$. We verify this in our simulations in the 
bottom left panels of Figs.~\ref{fig:intro}, \ref{fig:intro3d} for two and three dimensions, respectively. In the ordered phase 
$(\tau<0)$, the binary mixture eventually fully phase separates.
We can also see this in our simulations in the top left panels of Figs.~\ref{fig:intro}, \ref{fig:intro3d}.

In the presence of a drive, novel features emerge. In particular, our model exhibits behavior reminiscent of systems with modulated phases in 
\textit{equilibrium}, with some important differences. A modulated phase at equilibrium would have $S_{\psi}(\mathbf{q})$ with a maximum at a 
nonzero wavenumber: $S_{\psi}(\mathbf{q}) \propto [\tau+\kappa( |\mathbf{q}|-q^*)^2]^{-1}$, with $q^*=2\pi/\lambda^*$ and $\lambda^*$ 
the characteristic wavelength of the spatial modulation.   In a scattering experiment, we would expect a large contribution at 
this wavenumber for any direction $\hat{\mathbf{q}}$. Conversely, in our model,  the applied drive breaks the rotational symmetry 
of our system and the  static structure factor $S_{\psi}(\mathbf{q})$ only has peaks along the drive direction, with 
$S_{\psi}(\mathbf{q}) \propto [\tau+\kappa_{\perp}|\mathbf{q}_{\perp}|^2+\kappa_{\parallel}(|q_{\parallel}|-q_{\parallel}^*)^2]^{-1}$, 
where $q_{\parallel }=\mathbf{q} \cdot \hat{\mathbf{x}}_{\parallel}$ is the component of the wavevector parallel to and 
$\mathbf{q}_{\perp}$ the component perpendicular to the drive direction $\hat{\mathbf{x}}_{\parallel}$. Here we use simulations and 
a coarse-grained field-theoretic approach to see how such a peculiar ``microemulsion'' phase may  develop. 
 
Another important limit worth mentioning is the out-of-equilibrium, low particle density $\bar{\rho}$ limit at non-zero drive. Here, we expect the inter-species repulsive interactions to be negligible and the lattice gas should reduce to the low density phase of 
a non-interacting driven diffusive system (DDS) \cite{AnomalousDiff,KLSFT3,DDSbook}. The excluded volume condition, however, should still be relevant 
and appears as an interaction ``along the drive" (see the term proportional to $g$ below).  The local particle density
$\rho \equiv \rho (\mathbf{r},t)=\rho_A(\mathbf{r},t)+\rho_B(\mathbf{r},t)$ would obey, in the frame moving with the mean particle 
velocity induced by the drive $\delta$, the Langevin equation
\begin{align}
\partial_t \rho=D[c \partial^2_{\parallel}+\nabla^2_{\perp}]\rho +g\partial_{\parallel}[\rho^2] +\xi_{\rho}(\mathbf{r},t), \label{eq:introDDS}
\end{align}
where $\xi_{\rho}(\mathbf{r},t)$ is a conserved noise satisfying 
$\langle\xi_{\rho}(\mathbf{r},t)\xi_{\rho}(\mathbf{r}',t')\rangle =-2D  ( \tilde{c} \partial^2_{\parallel}  +\nabla_{\perp}^2)\delta(\mathbf{r}-\mathbf{r}')\delta(t-t')$, 
and  $c,\tilde{c}$  are coefficients reflecting the anisotropy in  the diffusion and noise, respectively. 
Detailed balance is violated and a renormalization group analysis leads to anomalous diffusion (only in the direction of the drive) for $d \le 2$
\cite{KLSFT3,AnomalousDiff}. 
We shall see that Eq.~\eqref{eq:introDDS} represents the coarse-grained dynamics of our model 
in the important limiting case of low particle density and no repulsive interaction between particle species. 
As in DDS, Eq.~\eqref{eq:introDDS} embodies a discontinuity singularity in the static structure factor $S(\mathbf{q})$, 
induced by detailed balance violation and the drive (see, e.g., Ref. \cite{DDSbook}). 
Note, however, that it has no characteristic peak of the kind shown in Figs.~\ref{fig:intro}, \ref{fig:intro3d} 
(see insets in bottom right panels) \cite{KLSFT1,KLSFT2}. 
Moreover, it is known that single-species systems with attractive interactions do not form modulated phases, either. 
If any spatial structures emerge (e.g., stripes of high and low densities), they are invariably aligned \textit{parallel} to the drive 
\cite{KLS1,KLS2,DDSbook}. 
Thus, the behavior analyzed in the two-species model presented here is remarkable and unintuitive, 
especially in light of the limiting, well-studied cases described above.
  
The remainder of this paper is organized as follows: In the next section we define the model. 
Then, in Sec.~\ref{sec:phenom}, we present simulation results on the disordered phase, including static structure factor 
calculations and  the characteristic wavevector $q_{\parallel}^*$ along the drive direction as a function of the drive $\delta$. 
In Sec.~IV, we derive a field theory corresponding to the lattice gas rules and show that, at the mean-field (Gaussian) level, it predicts 
some features of the disordered phase, yet missing the most prominent properties. Fluctuation corrections are discussed in Sec.~V, and we show 
that a simple perturbative approach is able to capture, if only qualitatively, the essential behavior exhibited by our driven lattice gas. 
We conclude in Sec.~VI and offer some directions for future investigation.

 \section{Model \label{sec:model}}
 
\begin{figure}[htp]
\centering
\includegraphics[width=0.3\textwidth]{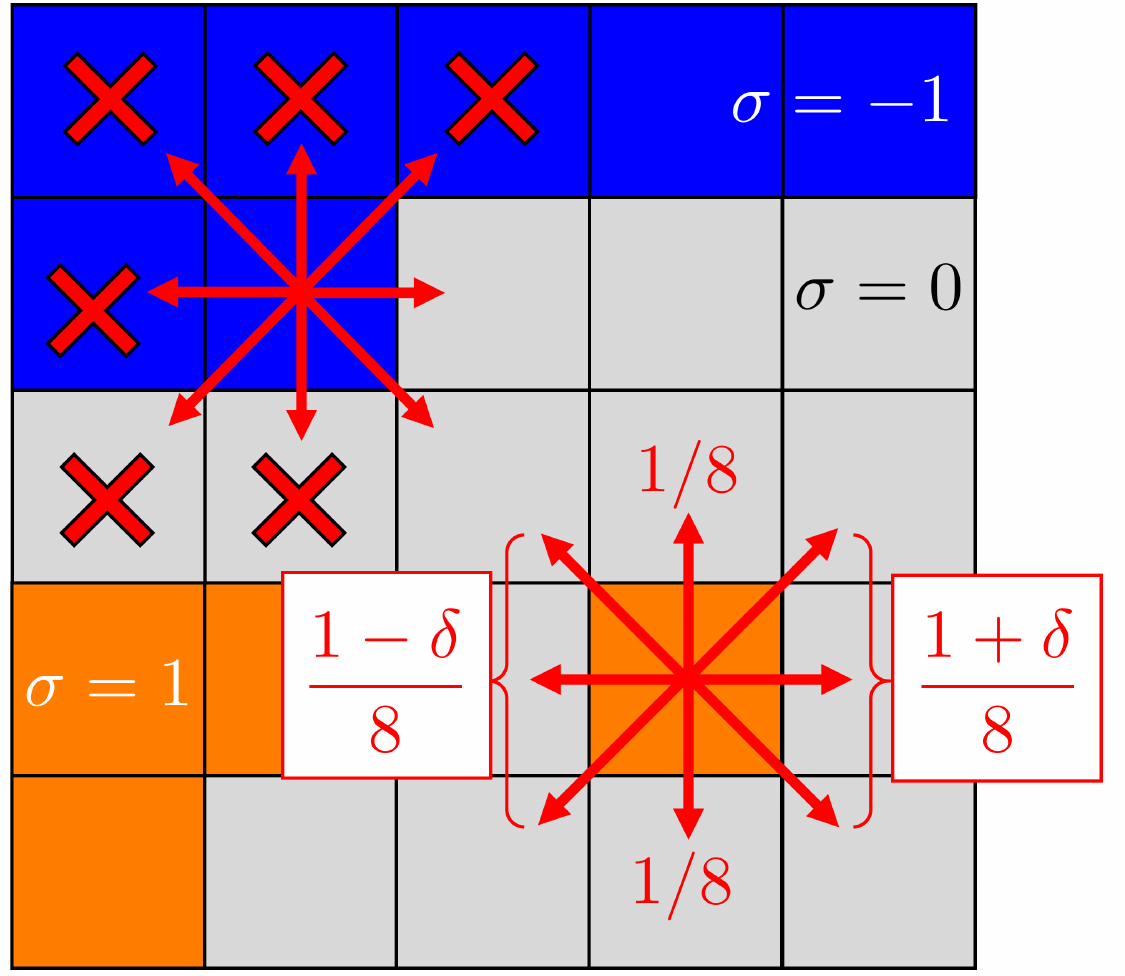}
\caption{\label{fig:setup} Update rules for the DWRLG shown on a segment of a square lattice. Orange $(\sigma=1)$ and blue $(\sigma=-1$) particles can hop in one of eight directions with the indicated probabilities. A particle hops to the target site as long as it is unoccupied and is not a nearest neighbor of a particle of the opposite type. Some prohibited hops (red x's) are shown for a blue particle. The generalization to the cubic lattice is straightforward: there are then six nearest neighbors and twelve next-nearest neighbor sites to which particles can hop. } 
\end{figure}

We consider a lattice gas with two particle species, $A$ (orange) and $B$ (blue), that occupy sites $\mathbf{x}=\ell(i,j)$ on a square lattice or $\mathbf{x}=\ell(i,j,k)$ on a cubic lattice (with $i,j,k$ integers), with $\ell$ the lattice spacing and $L$ the length of the lattice. The space of allowed configurations is defined by the restrictions that (1) all particles
occupy distinct sites, and (2) $A$-$B$ nearest-neighbor pairs are prohibited, as shown in  Fig.~\ref{fig:setup}.
We impose periodic boundary conditions in all directions. 
It is convenient to introduce spin variables $\sigma_{\mathbf{x}}=0,\pm1$ at each site, defined so:
\begin{equation}
\sigma_{\mathbf{x}} = \begin{cases}
1, & \mathbf{x}~\mbox{ is occupied by $A$} \\ -1, & \mathbf{x}~\mbox{ is occupied by $B$} \\ 0, & \mathbf{x}~\mbox{ is empty}.
\end{cases}. \label{eq:spindef}
\end{equation}
   The numbers of A particles $N_{A}$, and of $B$ particles $N_B$ remain fixed, as do the average densities, $\bar{\rho}_{A,B} = N_{A,B}\ell^d/L^d$. 

    The hopping rules are those in Ref.~\cite{DRWLG1} with $a=1/4$: At each time step, we pick a random particle at location $\mathbf{x}$ and move it to 
    a nearest-neighbor (NN) or next-nearest-neighbor (NNN) site $\mathbf{x}+\Delta\mathbf{x}$ with the following probabilities:\begin{equation}
\omega_{\mathbf{x}\rightarrow\mathbf{x}+\Delta \mathbf{x}}=\frac{1}{N_n }\begin{cases}
1+\delta, & \Delta\mathbf{x}\cdot \hat{\mathbf{x}}_{\parallel} >0 \\
1-\delta, & \Delta\mathbf{x}\cdot \hat{\mathbf{x}}_{\parallel}<0 \\
1, & \Delta\mathbf{x}\cdot \hat{\mathbf{x}}_{\parallel} = 0
\end{cases}, \label{eq:latticeupdates}
\end{equation}
where $0\leq \delta\leq 1$ is the drive strength, $\hat{\mathbf{x}}_{\parallel}$ is the drive direction, and $N_n$ is the number of NN and NNN sites. ($N_n=8$ and 18  the square and simple cubic lattices, respectively.) The particle displacement is accepted subject to the restrictions mentioned above. Note that the rules are completely symmetric for the two species, so there is an ``Ising-like'' symmetry $\sigma_{\mathbf{x}} \rightarrow -\sigma_{\mathbf{x}}$  (so long as we keep the number of particles of each type the same, with $N_A=N_B$). After each such update, time advances by $1/N$, where $N=N_A + N_B$ is the total number of particles.  We define one Monte-Carlo step (MCS) as having completed $N$ such attempts and will label MCS by $n=1,2,...,N_{\mathrm{MCS}}$. In our simulations, the runs involve typically $N_{\mathrm{MCS}}$ up to $5 \times 10^7$, with the longest runs going up to $10^8$ MCS. Measurements are taken typically after discarding the initial $4 \times 10^6$ MCS so that the system has arrived at a (reasonable) stationary state.

 In equilibrium (no drive, $\delta=0$), the model exhibits a phase transition at a critical density $\bar{\rho}^* = 0.617(1)$ and $\bar{\rho}^*=0.3543(1)$ in two and three dimensions, respectively \cite{eqRWLG1}. The transition is in the Ising universality class and describes the usual demixing transition of a binary mixture at sufficiently high densities. Examples of the disordered and ordered phases are shown in the left columns of Figs.~\ref{fig:intro} and \ref{fig:intro3d}. One may track the transition via the structure factor associated with the ``charge'' order parameter field $\psi(\mathbf{r})$:
\[
S_{\psi} (\mathbf{q})=\frac{1}{L^d}\sum_{\mathbf{x},\mathbf{r}} \langle \sigma_{\mathbf{x}} \sigma_{\mathbf{x}+\mathbf{r}} \rangle e^{i\mathbf{q} \cdot \mathbf{r}}=  \frac{\langle | \sigma(\mathbf{q})|^2 \rangle}{L^d},
\]
where $\sigma(\mathbf{q})$ is the Fourier-transformed particle configuration $\sigma_{\mathbf{x}}$ and $L^d$\ the volume of the square $(d=2)$ or cubic $(d=3)$ lattice. For a phase with a characteristic length, we expect to find peaks in $S_{\psi}(\mathbf{q})$. We will analyze the formation of this ``microemulsion peak'' using simulation and a field-theoretic approach.

\section{\label{sec:phenom} Microemulsions in Simulations}

\begin{figure}[htp]
\centering
\includegraphics[width=0.45\textwidth]{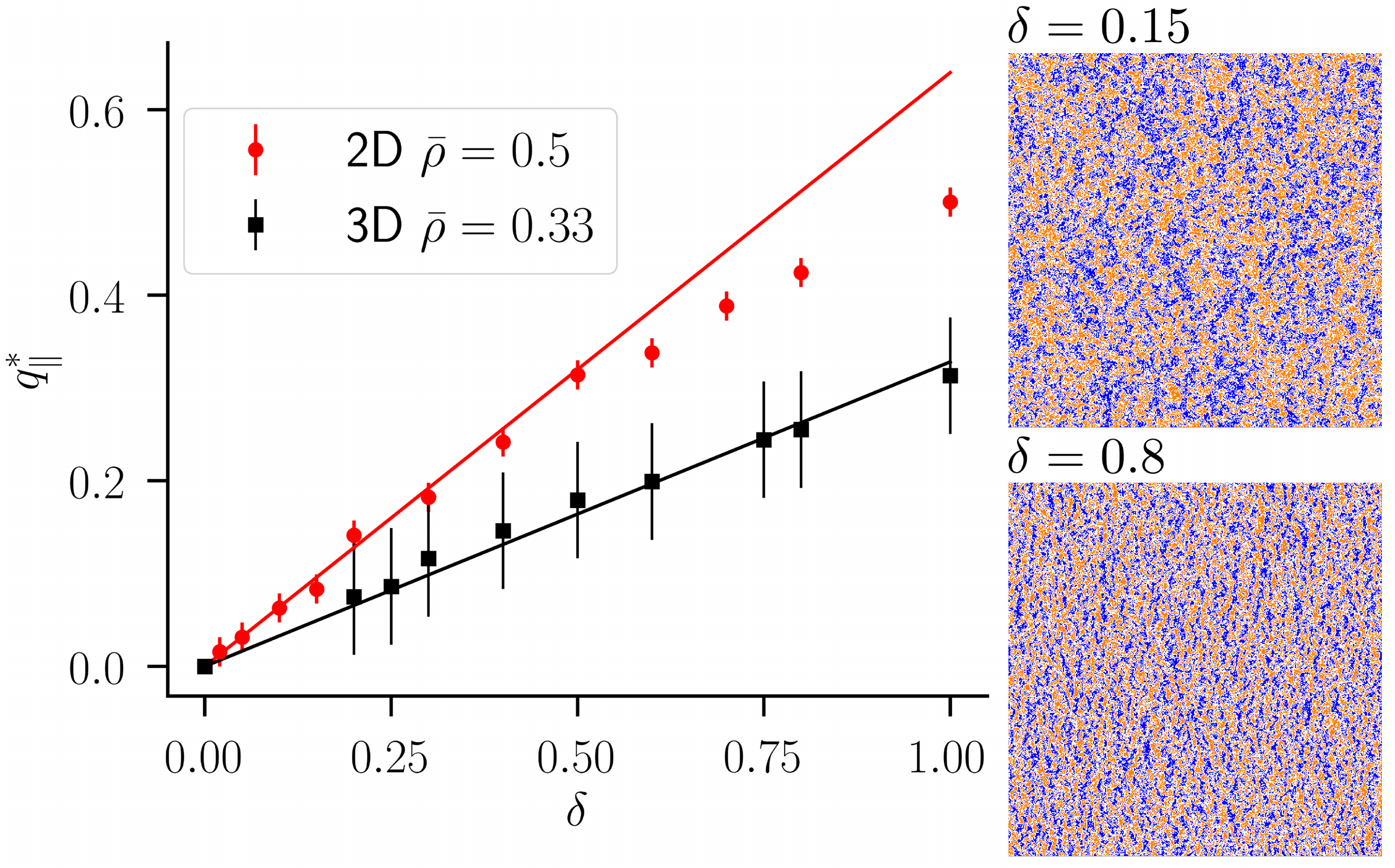}
\caption{\label{fig:stripesize}  Structure factor peak position $q_{\parallel}^*$ for two- and three-dimensional systems in the microemulsion phase. Note that the peak varies linearly with the drive $\delta$ for small drives: $q_{\parallel}^* \propto \delta$. The error bars indicate $2\pi/L$ which sets the $\mathbf{k}$-space resolution $(L=400$ for 2D and $L=100$ for 3D). This  linear variation of $q_{\parallel}^*$ with $\delta$ in the disordered phase is consistent with our field-theoretic analysis. Typical system snapshots are shown for the square lattice on the right panels for $\bar{\rho}=0.5$ and the indicated drive values $\delta$. } 
\end{figure}

We begin with simulation results that give the basic phenomenology of this system. We have simulated both two- and three-dimensional systems, but since it is easier to go to larger system sizes in two dimensions, this will be our primary focus.  We will also primarily focus on the behavior of the system below the transition point $(\bar{\rho}<\bar{\rho}^*)$ where we  find a disordered ``microemulsion'' phase. As discussed in the previous section, although we do not have the distinct stripe order at low densities, the small fluctuating domains of particles in the system are characterized by a distinct length scale along the drive direction. This manifests as a peak in the structure factor $S_{\psi}(\mathbf{q})$ at $\mathbf{q}=(q_{\parallel}^*,0)$. As one approaches the critical density $\bar{\rho} \rightarrow \bar{\rho}^*$, this peak value begins to diverge with the system size, as we develop the uniform stripe order seen in the top right panels of Figs.~\ref{fig:intro},\ref{fig:intro3d}.
For $\bar{\rho}< \bar{\rho}_*$, the position of the peak $q_{\parallel}^*$ in the drive direction grows linearly with the drive $\delta$ in both two and three dimensions, as shown in Fig.~\ref{fig:stripesize}.
 This is also reflected in the coarser texture of the microphase pattern for smaller $\delta$, evident in the configuration snapshots.

\begin{figure}[htp]
\centering
\includegraphics[width=0.4\textwidth]{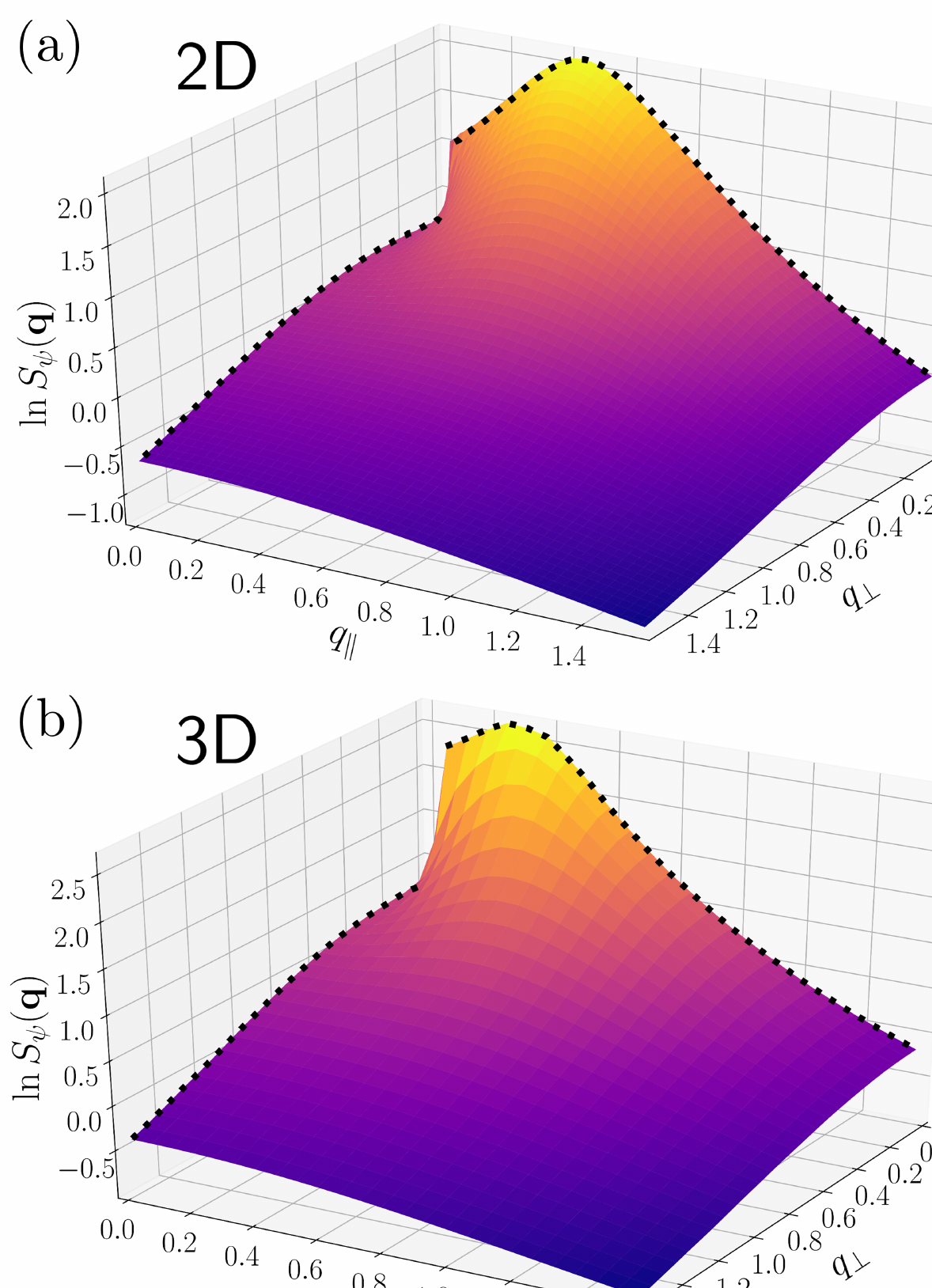}
\caption{\label{fig:Skfull} Static structure factor $S_{\psi}(\mathbf{q})$ calculated from simulations for a drive $\delta=0.8$ in the 
low-density phase  in (a) two dimensions with $\bar{\rho}=0.5$ ($L^2$ sites with $L=400$) 
and (b) three dimensions, with $\bar{\rho}=0.33$ ($L^3$ sites with $L=100$). Note the jump discontinuity at the origin, characteristic of a driven diffusive system. The black dashed lines show the limiting forms of  $S_{\psi}(\mathbf{q})$ for $\mathbf{q}=(q_{\parallel},q_{\perp}=0)$ and $\mathbf{q}=(q_{\parallel}=0,q_{\perp})$. The  $\mathbf{q}=(q_{\parallel}=0,q_{\perp})$ direction has  a smooth, featureless, monotonically-decreasing shape for the structure factor, while the other direction $\mathbf{q}=(q_{\parallel},q_{\perp}=0)$ has a peak at a non-zero characteristic wavenumber $q_{\parallel}^*$.} 
\end{figure}

\begin{figure}[htp]
\centering
\includegraphics[width=0.4\textwidth]{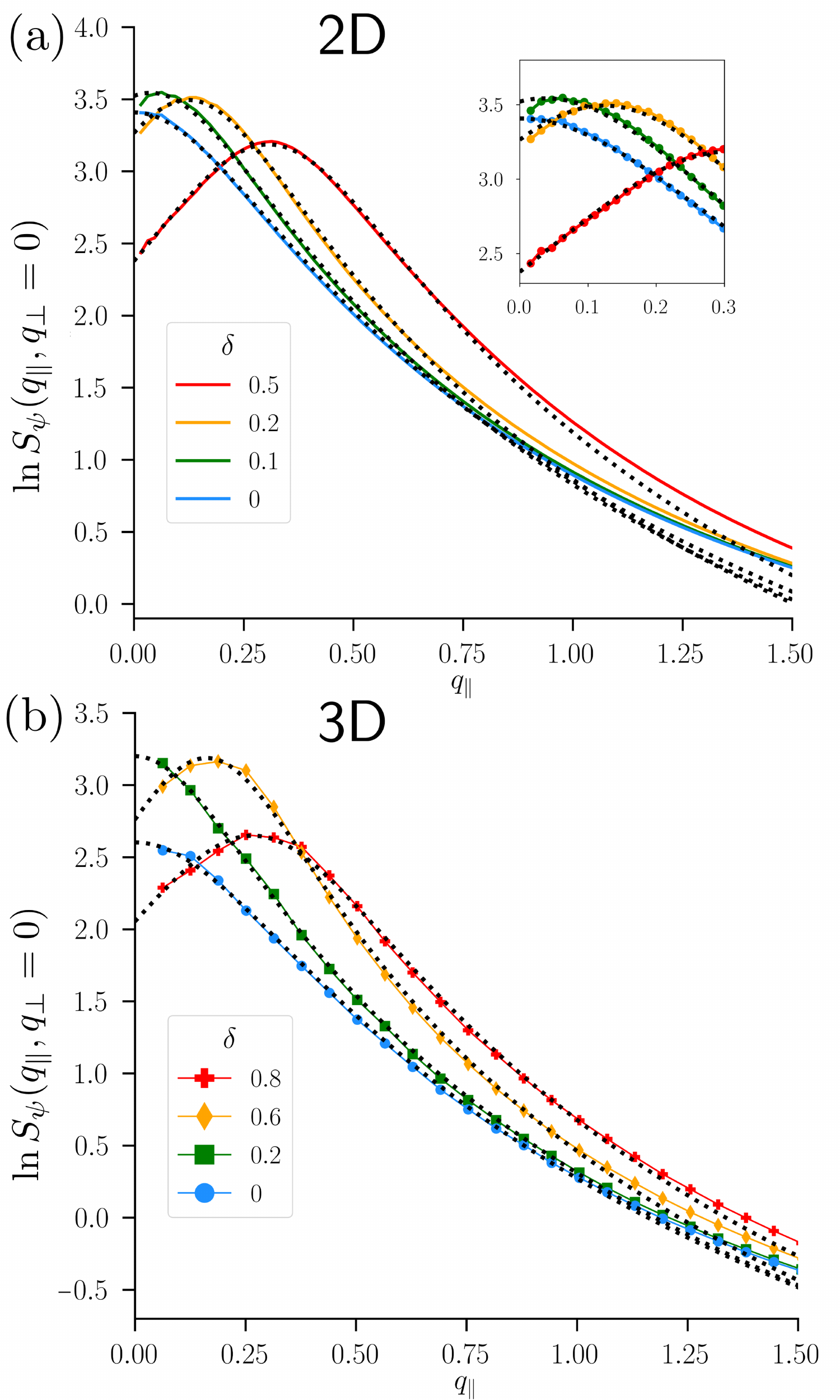}
\caption{\label{fig:sfactor}  Static structure factor at different values of the drive $\delta$ for the charge field $\psi=\rho_A-\rho_B$ as a function of $q_{\parallel}$ for $|\mathbf{q}_{\perp}|=0$ in two- (a) and three- (b) dimensional systems with $L=400$ and $L=100$, respectively. In both cases, we are below the phase separation transition and have total particle densities $\bar{\rho}=0.5$ in (a) and $\bar{\rho}=0.33$ in (b).   The dotted lines are fits to the form given by Eq.~\eqref{eq:skphenom} (with three fitting parameters $\nu_{\parallel}$, $\tau_{\parallel}$, and $q_{\parallel}^*$).   Because this form is only appropriate for small $q$ where the details of the lattice structure are not important, we fit for data points between $0<q_{\parallel} \lesssim0.8$.   Note that the structure factor develops a peak at a non-zero $q_{\parallel} $ when $\delta>0$. The inset in (a) has the same axes as the main plot and shows the structure factor shape near the origin where one finds a linear \textit{kink}.} 
\end{figure}

In the previous work on this model \cite{DRWLG1}, the conjectured simplest form of the static structure factor $S_{\psi} \equiv S_{\psi}(\mathbf{q})$, consistent with a dominant contribution at a nonzero $\mathbf{q}=(q_{\parallel}^*,0)$, reads
 \begin{widetext}\[
S_{\psi} [\mathbf{q}=(q_{\parallel},\mathbf{q}_{\perp})]= \frac{\nu_{\parallel}^{}q_{\parallel}^2+\nu_{\perp}^{} |\mathbf{q}_{\perp}|^2   +\ldots}{  \tau_{\parallel}^{}q_{\parallel}^2+ \tau_{\perp}^{}|\mathbf{q}_{\perp}|^2   +q_{\parallel}^2 (|q_{\parallel}|-q_{\parallel}^*)^2+\gamma_{\times}^{}q_{\parallel}^2 |\mathbf{q}_{\perp}|^2+\gamma_{\perp}^{}|\mathbf{q}_{\perp}|^4+\ldots} , \label{eq:skphenom}
\] \end{widetext}
with ellipses indicating corrections in higher powers of the momentum $\mathbf{q}$.
The new interesting feature here is the cubic term,  $-2q_{\parallel}^* |q_{\parallel}|^3$,  in the denominator, which is  responsible for the peak in the structure factor.  The cubic term is  unexpected from a mean field analysis and a naive continuum limit of the model, which yields only even powers of $q_{\parallel}$ and no drive-dependence in any of the terms.\footnote{Also, the two-point correlation function must obey an underlying parity symmetry, which constrains it to be symmetric under $q_{\parallel} \rightarrow -q_{\parallel}$. The drive breaks this symmetry, but this feature can only appear at the level of three-point (or larger) correlation functions.}  The cubic term also yields a   linear kink   in the static structure factor near the origin along the drive direction, as evidenced in Figs.~\ref{fig:Skfull} and \ref{fig:sfactor}, where we fit the conjectured form in Eq.~\eqref{eq:skphenom} to simulation data for $\mathbf{q}_{\perp}=0$.    Note how well the form reproduces the features at small $q_{\parallel}$. In two dimensions, we have verified that the kinked form is not a finite-size effect by checking that the form does not change as we vary the lattice size: $L=100,200, 400$. Also, as far as can be estimated from simulations, the peak exists for any $\delta>0$. Our objective, then, is to explore how such a structure factor form may come about from a field-theoretic analysis of this model.

The static structure factors in two and three dimensions as a function of $\mathbf{q}$, both parallel and perpendicular to the drive, are shown in Fig.~\ref{fig:Skfull}, as measured from simulations. It is  worth noting that the static structure factors exhibit a characteristic jump discontinuity near the origin, captured by our conjectured form in Eq.~\eqref{eq:skphenom} by setting $\nu_{\parallel}/\tau_{\parallel} \neq \nu_{\perp}/ \tau_{\perp}$.  We also see that the structure factors monotonically decrease along the direction $q_{\perp}$ perpendicular to the drive for $q_{\parallel}=0$.
Along this direction, the structure factor is well-approximated by  a simple  Ornstein-Zernike form, corresponding to the $q_{\parallel}=0$ case in Eq.~\eqref{eq:skphenom}.

\begin{figure}[htp]
\centering
\includegraphics[width=0.5\textwidth]{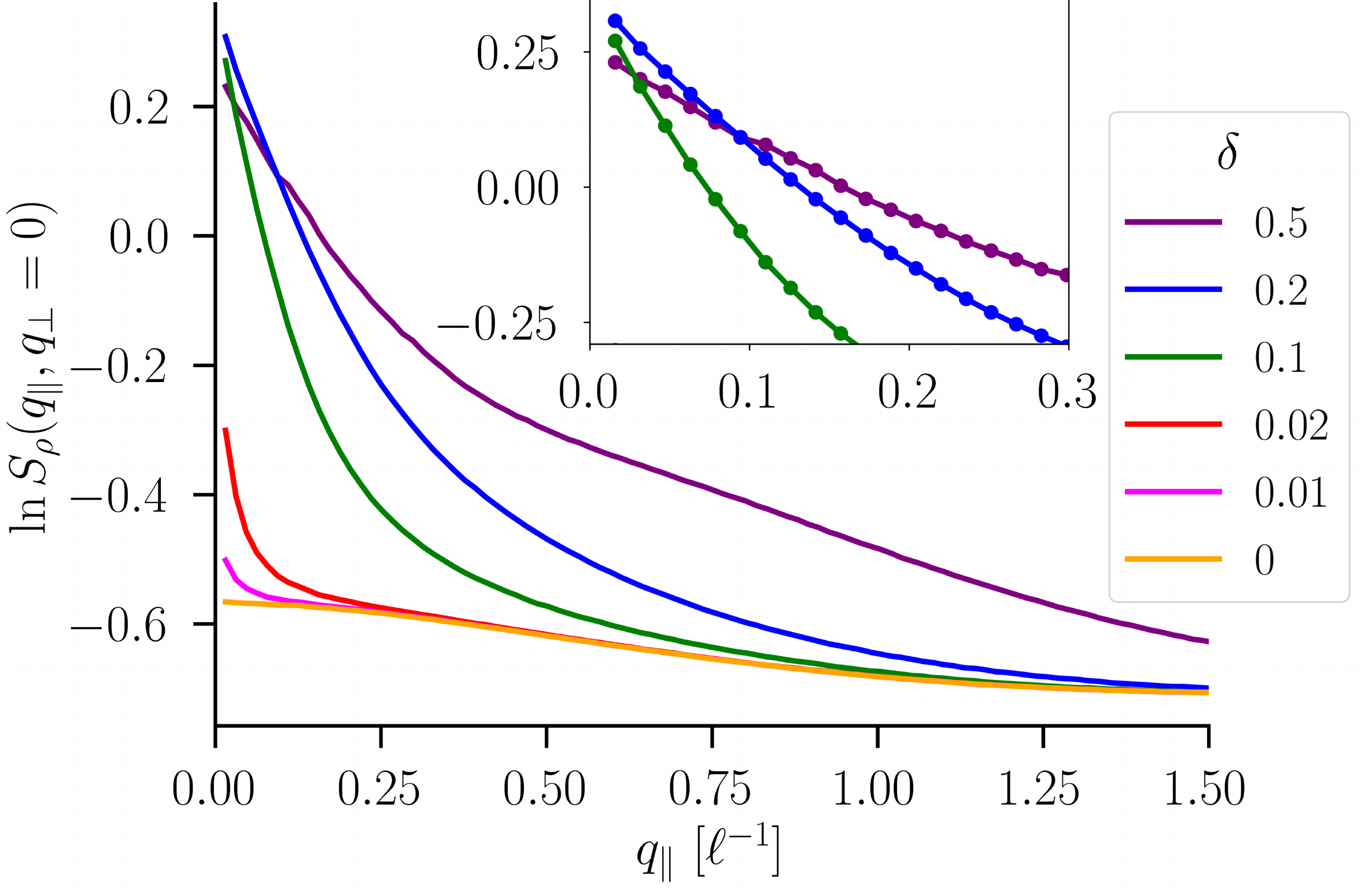}
\caption{\label{fig:densfactor}  Static structure factor $S_{\rho}(\mathbf{q})$ [with $\mathbf{q}=(q_{\parallel},q_{\perp})$] of the density field $\rho=\rho_A+\rho_B$,
 as a function of $q_{\parallel}$, with $q_{\perp}=0$, measured in simulations for drives $\delta$ as indicated.  $S_{\rho}$ is calculated using a two-dimensional system with $L^2=400^2$ lattice sites at an average particle density of $\bar{\rho}=0.5$ (equal fractions of\ $A$ and $B$ particles). The inset, which has the same axes as the main plot, shows a detail
 of the small-$q_{\parallel}$ region. $S_\rho$ has a cusped, sharp increase near $q_{\parallel}=0$, consistent with a logarithmic divergence. Unlike in the case of the charge structure factors $S_{\psi}$ shown in  Fig.~\ref{fig:sfactor}(a), we do not find a peak here.} 
\end{figure}

 Before moving on, let us consider the static structure factor $S_{\rho}(\mathbf{q})$ for the density field $\rho \equiv \rho_A+\rho_B$ in two dimensions. This also has an interesting structure when the drive $\delta$ is applied, as can be seen in Fig.~\ref{fig:densfactor}. Note how there is  a kinked structure near $q_{\parallel}=0$ that is of the \textit{opposite sign} as $S_{\psi}(\mathbf{q})$ in Fig.~\ref{fig:sfactor}.     
  As shown in Fig.~\ref{fig:densfactor}, the kinked increase at $q_{\parallel}=0$ grows with increasing drive $\delta$ but starts to decrease for the largest drives. We will show that the kink near the origin likely comes from the fluctuation corrections in the field-theoretic description, just as for the charge fields.  We should remark that $S_{\rho}(\mathbf{q})$ does not appear to exhibit any discernible signatures of the characteristic wavenumber $q_{\parallel}^*$. On closer examination, however, we can detect a shoulder in the $\delta=0.5$ data, at roughly $2q_{\parallel}^*$. We suspect that, especially for the small $\delta$ systems, the enhancement manifest in $S_{\rho}$ is mostly shrouded by the much larger effect near $q_{\parallel}=0$ here. These properties are only true at small $\bar{\rho}$. In the high density, ordered phase (with $\bar{\rho}> \bar{\rho}^*$), $S_{\rho}(\mathbf{q})$ displays a peak at $\mathbf{q}=(q_{\parallel}=2q_{\parallel}^*,q_{\perp}=0)$, because high density stripes of $A$ and $B$ particles alternate, separated by low density regions (of many vacancies) \cite{DRWLG1}.

\begin{figure}[htp]
\centering
\includegraphics[width=0.34\textwidth]{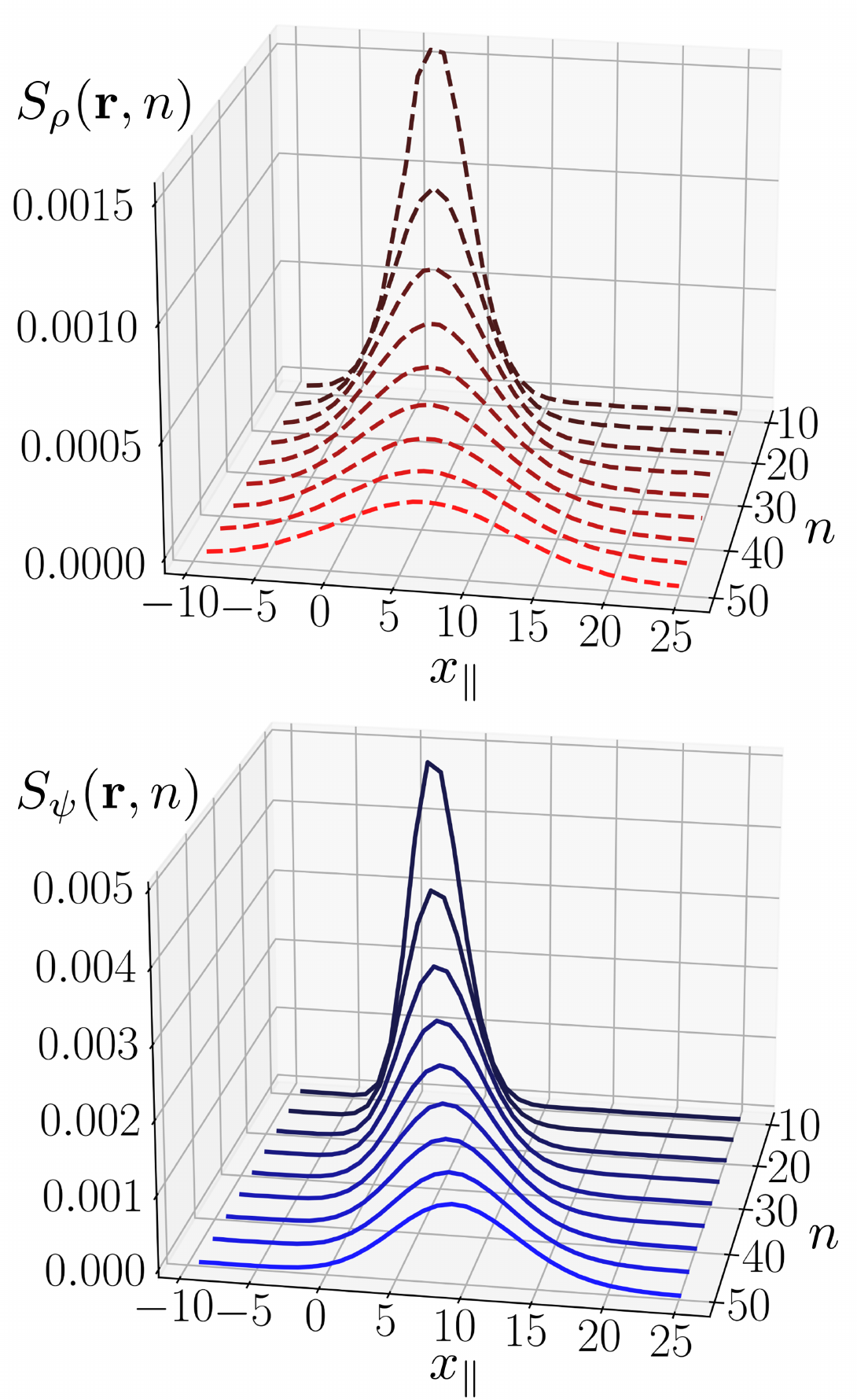}
\caption{\label{fig:Sxt} Dynamic  (real-space) structure factors [defined in Eqs.~(\ref{eq:sxt1},\ref{eq:sxt2})] calculated from a two-dimensional simulation with equal fractions of $A$ and $B$ particles with $\bar{\rho}=0.1$ and $\delta=0.3$ (system size $L=50$). The factors are evaluated along the drive direction $\mathbf{r}=(x_{\parallel},0)$.  The charge field structure factor $S_{\psi}(\mathbf{r},n)$ (lower panel) propagates more rapidly and spreads more slowly than the density field structure factor, $S_{\rho}(\mathbf{r},n)$ (upper panel). Both have Gaussian shapes. These properties can be understood within a field-theoretic framework. } 
\end{figure}

 Of course, the static structure factors $S_{\psi,\rho}(\mathbf{q})$ display only part of the interesting behavior we observe. The dynamic structure factors $S_{\psi,\rho}(\mathbf{r},n)$ for the charge and density fields are also important, as they directly capture not only the diffusive properties of the particles, but also the advection. In particular, we may measure the unequal time correlations given by
\begin{eqnarray}
S_{\psi }(\mathbf{r},n)& = & \frac{1}{L^d}\sum_{\mathbf{x},t} \langle \sigma_{\mathbf{x}}(t) \sigma_{\mathbf{x}+\mathbf{r}} (t+n)\rangle  \label{eq:sxt1} \\
S_{\rho} (\mathbf{r},n) & = & \frac{1}{L^d}\sum_{\mathbf{x},t} [\langle|\sigma_{\mathbf{x}}(t)|| \sigma_{\mathbf{x}+\mathbf{r}} (t+n)| \rangle-\bar{\rho}^2], \label{eq:sxt2}
\end{eqnarray}
with $\sigma_{\mathbf{x}}(t)$ the spin configuration at time step $t$ and lattice site $\mathbf{x}$ [see Eq.~\eqref{eq:spindef}], and $\bar{\rho}$ the average particle density in the system (total number of $A$ and $B$ particles divided by the total number of lattice sites).
The angular brackets $\langle \ldots \rangle$ denote an average over many simulation runs. The results for the low density phase  in two dimensions are shown in Fig.~\ref{fig:Sxt}.  

Note that the charge and density dynamic structure factors have very different behaviors: The peaks of the charge field structure factor $S_{\psi}(\mathbf{r},n)$ broaden more slowly and move more rapidly compared to the density field structure factor $S_{\rho}(\mathbf{r},n)$. This is consistent with a larger characteristic velocity and a smaller diffusivity for the charge field. Moreover, the shape of these structure factors is approximately Gaussian. We shall justify this particular form and derive these properties in the following section when we explore the field-theoretic formulation of this model.

\section{Coarse-grained field-theoretic model\label{sec:FTsetup}}
 
To understand how these unusual properties arise from being driven, even if only qualitatively, we will attempt an approach based on coarse-graining 
our discrete lattice system to a continuum field-theory. Starting with a set of stochastic update rules, such as those described by 
Eq.~\eqref{eq:latticeupdates}, it is straightforward to derive a master equation for the evolution of the probability 
$P(\{ \sigma_{\mathbf{x}} \},t)$ of observing a particular lattice configuration $\{ \sigma_{\mathbf{x}} \}$ at time $t$. 
From here, there are standard procedures for converting that description into one based on a Langevin equation for 
coarse-grained density fields $\rho_{A,B}(\mathbf{x},t)$. A common approach employs the Martin-Siggia-Rose-Janssen-de~Dominicis (MSRJD) 
formalism \cite{MSR,Janssen,deDominicis} which translates these equations into a field-theoretic, ``dynamical action'' $\mathcal{J}$ 
involving both the $\rho_{A,B}$'s and corresponding ``response fields'' $\hat{\rho}_{A,B}(\mathbf{x},t)$. Following the general procedure 
for particle hopping models \cite{BiroliDoi}, the dynamical action reads
\begin{align}
\mathcal{J}[\rho_{A,B},\hat{\rho}_{A,B}] &=\sum_{\alpha=A,B}\int \mathrm{d}t \bigg\{ \sum_{\mathbf{x} } \hat{\rho}_{\alpha} \partial_t \rho_{\alpha}  \nonumber \\ & {} +\sum_{\langle \mathbf{x} \mathbf{y}\rangle}W^{\alpha}_{\mathbf{x}\rightarrow \mathbf{y}}\left[1-e^{\hat{\rho}_{\alpha}(\mathbf{x})-\hat{\rho}_{\alpha}(\mathbf{y})}\right]\bigg\},
\label{eq:action}
\end{align}
where $W^{\alpha}_{\mathbf{x}\rightarrow\mathbf{y}}$ are the hopping rates from site $\mathbf{x}$ to $\mathbf{y}$ for particles 
of type $\alpha=A,B$ [which can be gleaned from the $\omega$'s in Eq.~\eqref{eq:latticeupdates}]. These $W$'s encode both the hopping and 
the exclusion rules and must vanish for any prohibited hops (see Fig.~\ref{fig:setup}). We sum over all pairs of NN and NNN sites 
$\langle \mathbf{x} \mathbf{y}\rangle$.  From here, the average over many stochastic realizations of the particle system may be 
computed for any functional of the densities $\mathcal{O}(\{ \rho_{A,B} \})$ (``observables'') using the action $\mathcal{J}$ via:
\begin{equation}
\langle \mathcal{O} (\{ \rho_{A,B} \})\rangle =\frac{1}{\mathcal{Z}} \int \mathcal{D} \rho_{A,B} \, \mathcal{D}\hat{\rho}_{A,B} \mathcal{O}(\{ \rho_{A,B} \})e^{-\mathcal{J}}, \label{eq:averaging}
\end{equation}
with $\mathcal{Z}$ providing normalization. Also, note that the Euler-Lagrange equations associated with extremizing $\mathcal{J}$
(with an appropriate interpretation of the response fields $\hat{\rho}_{A,B}$) lead us to Langevin equations for the densities $\rho_{A,B}(\mathbf{x},t)$. As these are more easily grasped intuitively, we will present 
the field theoretic formulation in such terms [see Eqs.~\eqref{eq:langevin} below].

A second method (Doi-Peliti \cite{Doi,Peliti}) of deriving a coarse-grained description of the dynamics and the dynamical action $\mathcal{J}$
involves reformulating the master equation for $P(\{ \sigma_{\mathbf{x}} \},t)$ using a Fock space where the probability distribution 
$P$ is encoded in a multi-particle state. Then, using a coherent state path integral representation of the master equation, one is able 
to derive a dynamical action of the same form as Eq.~\eqref{eq:action}. In principle, these two methods yield equivalent field theories 
\cite{BiroliDoi}, but the mapping is non-trivial. Also, a special difficulty in our case is that both the excluded volume constraint
and the $A,B$ particle next-nearest neighbor exclusion rule are not easily incorporated. One possibility, introduced by 
van Wijland \cite{vanwijland}, is to apply the exclusion rules at the level of the master equation, which is exact but difficult 
to interpret when we coarse-grain. A review can be found in, e.g., Ref.~\cite{RReview}. 
This procedure is quite involved and some details are provided in Appendix~\ref{appx:doipeliti} for the interested reader. 

Encouragingly, after taking the continuum limit and expanding around small density fluctuations, both the Doi-Peliti and MSRJD methods 
yield the same structure for the field theory, with minor variations in the dependencies of the coupling constants on the microscopic 
parameters $\bar{\rho}_{A,B}$ and $\delta$.  In the following we use results from the Doi-Peliti approach using van Wijland's method 
for excluded volume interactions.

In the paradigm of this coarse-grained continuum description, we assume the density fields $\rho_{\alpha}$ and their corresponding 
response fields $\hat{\rho}_{\alpha}$ are slowly varying in space and time ($\mathbf{r},t$). To construct a perturbation theory and compare to 
our simulation results, we transform to more convenient fields (i.e., ones which diagonalize the quadratic part in $\mathcal{J}$: 
the total density $\rho \equiv \rho_A+\rho_B$ and the ``charge'' order parameter $\psi \equiv \rho_A-\rho_B$). 
We have to expand the fields $\rho$ and $\psi$ around some uniform concentrations $\bar{\rho}$ and $\bar{\psi}$ which, 
along with the drive $\delta$, are our control parameters in the simulations. With the replacements 
$\rho = \bar{\rho}+ \phi_+$ and $\psi = \bar{\psi} +\phi_-$, we find that, to leading order in the fluctuations 
$\phi_{\pm}$ and the averages, the Langevin equations read
\begin{equation}
\begin{cases}\partial_t \phi_{+}=&D  _{+}\nabla^2 \phi_{+}-v_+  \partial_{\parallel} \phi_{+} - u_{+}\bar{\psi}\partial_{\parallel} \phi_{-}\\&{}+\frac{g_+}{2}\partial_{\parallel} (\phi_{+}^2)-\frac{g_-}{2}\partial_{\parallel}(\phi_{-}^2)+\xi_{+} \\
\partial_t \phi_{-}=&D _{-}\nabla^2 \phi_{-}-v_-\partial_{\parallel} \phi_{-}\\&{}+ u_{-}\bar{\psi}\partial_{\parallel} \phi_{+}+g_0\partial_{\parallel}(\phi_{-}\phi_{+})+\xi_{-}
 \end{cases} ,\label{eq:langevin}
\end{equation}
where the coefficients of all the terms carry appropriate units of space (the lattice spacing $\ell$) and time (MCS) which will be suppressed.
With this understanding, we find (the leading constants and lowest-order corrections in powers of  $\bar{\rho}$ and $\bar{\psi}$ for)
the diffusions $D_{ \pm} \approx 3 /8$ and 
the velocities $v_{\pm} = \bar{v} \pm v_d/2$ with average $\bar{v} = (v_+ + v_-)/2 \approx  \delta (3-22\bar{\rho})/4 $ 
and the difference $v_d = v_+ - v_- \approx -5 \bar{\rho} \delta /2$. 
The other velocities, $ u_{+}\approx 7 \delta/4$ and  $ u_{-}\approx 3 \delta/4 $, do not appear in this study, 
as we focus on neutral systems ($\bar{\psi}=0$) only.
While $\nabla^2=\nabla_{\perp}^2+\partial_{\parallel}^2$ represents isotropic
diffusion, there are  DDS-like anisotropic, 
non-linear couplings $g_{-}=  \sqrt{2}\delta $, $g_{+}=4 \sqrt{2} \delta $, and $g_{0}=3 \sqrt{2} \delta $ all proportional to the drive $\delta$. 
As we shall see, these interaction terms will generate anisotropy in the diffusion terms. We also find conserved noises for both 
the charge and density, with correlations given by: 
$\langle \xi_{+} \xi_{+} \rangle=-2N_{+} \bar{\rho}\nabla^2\delta(\mathbf{r}-\mathbf{r}')\delta(t-t')$ and 
$\langle \xi_{-}\xi_{-} \rangle=-2N_{-}\bar{\rho}\nabla^2\delta(\mathbf{r}-\mathbf{r}')\delta(t-t')$, with $N_{\pm}\approx 3/8$.
When $\bar{\psi}\neq 0$, there are nonzero cross-correlations, 
$\langle \xi_{+}\xi_{-}\rangle=-2N_{+-} \bar{\psi} \nabla^2 \delta(\mathbf{r}-\mathbf{r}')\delta(t-t')$, with $N_{+-} \approx 3/8$. 
While it is of course possible to obtain more detailed expressions for the coupling constants in terms of the microscopic 
control parameters $\bar{\rho}$, $\bar{\psi}$, and $\delta$, they are not needed for our objective of understanding the 
coarse-grained features of the dynamics at small particle density $\bar{\rho}$. 

Note that when either of the two species vanishes, so that $\rho=\psi=\rho_{A,B}$, both equations in Eq.~\eqref{eq:langevin} 
reduce to the DDS Langevin equation in Eq.~\eqref{eq:introDDS} with $g=g_{0}=g_{+}-g_{-}$ as the derivative coupling. 
This is an important limiting case as we would expect our model to reduce to the single-species DDS when we remove the 
repulsive interactions between particle species. The other important case is $\delta=0$, where we would expect phase separation 
of the $A$ and $B$ particles at sufficiently high densities. In this case, Eq.~\eqref{eq:langevin} reduces to a pair of independent diffusion equations. To understand the system at higher particle densities, higher-order terms in the fields $\phi_{\pm}$
need to be included. If we further assume the density fluctuations $\phi_+$ relax faster than the charge $\phi_-$ and integrate out the former, 
the effective equation for the latter reduces to an Ising system with conserved dynamics (model B). 
However, this is a crude approximation and a more careful analysis would involve accounting for both fields $\phi_{\pm}$ and 
considering the action at high particle densities $\bar{\rho}$. Such a line of pursuit is beyond the scope of this paper. 
Here we focus on the low-density phase, and defer a detailed analysis of phase separation to future work.

We now discuss the dynamic action associated with the Langevin equations,  Eqs.~(\ref{eq:langevin}). To compare with simulation results, 
we impose equal average densities ($\bar{\psi}=0$) and find that,  
up to cubic terms in the fields and leading order in spatial derivatives, the action reads 
\begin{eqnarray}
& & \mathcal{J} = \int \mathrm{d}\mathbf{r} \,  \mathrm{d}t\bigg\{ \sum_{a=\pm} \Big[\hat{\phi}_{a}(\partial_t -D_a  \nabla^2 +v_a\partial_{\parallel} )\phi_a   \nonumber  \\
& & \qquad \quad{} +\bar{\rho} N_a\hat{\phi}_a \nabla^2 \hat{\phi}_a\Big]+g_0\phi_-\phi_+\partial_{\parallel}\hat{\phi}_-+\frac{g_+}{2} \phi_+^2 \partial_{\parallel} \hat{\phi}_+ \nonumber \\ & & \qquad \quad  {}-\frac{g _-}{2}\phi_-^2\partial_{\parallel} \hat{\phi}_+\bigg\}. \label{eq:actionDRWLG}
\end{eqnarray}
In the disordered phase, $\bar{\rho}<\bar{\rho}^*$, the diffusion constants $D_{\pm}$ are both positive. A scaling analysis 
(taking $\mathbf{r} \rightarrow \Lambda \mathbf{r})$ shows that  the scaling dimension of the drive couplings $g_{0,\pm}$ is $1-d/2$, 
so that the upper critical dimension is $d_c=2$ for the $\bar{\rho}<\bar{\rho}^*$ regime, as it is for the single-species DDS at low densities 
\cite{KLSFT3}. We therefore conclude that higher-derivative coupling terms are irrelevant in the renormalization group sense. 

Next, we set up a diagrammatic expansion using the action of Eq.~\eqref{eq:actionDRWLG}. 
There are two kinds of propagators generated by the quadratic terms in the fields $\phi_{\pm}$ and $\hat{\phi}_{\pm}$.
In the Fourier domain, we have the (bare) correlation functions 
$ \langle\phi_{\pm}( \mathbf{q},\omega)\phi_{\pm}(\mathbf{q}',\omega')  \rangle \equiv 
C_{\pm} (\mathbf{q},\omega)\delta(\mathbf{q}+\mathbf{q}')\delta(\omega'+\omega)$ 
and the response functions 
$\langle \hat{\phi}_{\pm}(\mathbf{q},\omega)\phi_{\pm}(\mathbf{q}',\omega') \rangle \equiv 
G_{\pm}(\mathbf{q},\omega)\delta(\mathbf{q}+\mathbf{q}')\delta(\omega'+\omega)$.
Using dotted and solid lines for the density and charge fluctuations, respectively, we denote these by the following:
\begin{equation}
\begin{cases} G_{\pm} = \dfrac{1 }{-i\left[\omega-v_{\pm}     q_{\parallel} \right]+D_{\pm} |\mathbf{q}|^2 } & \parbox{1.4cm}{\begin{fmfgraph*}(43,20)
\fmfleft{t1,v1,v3,t2} \fmfright{t3,v2,v4,t4} \fmfset{dot_len}{1mm} \fmf{plain_arrow,tension=1.5}{v1,v2}     \fmf{dots_arrow,tension=1.5}{v3,v4}   
 \end{fmfgraph*}} \\[10pt]
C_{\pm}   = \dfrac{2N_{\pm}   \bar{\rho}  |\mathbf{q}|^2 }{\left[\omega- v_{\pm}  q_{\parallel}\right]^2+\left[D_{\pm} |\mathbf{q}|^2 \right]^2} & \parbox{1.4cm}{\begin{fmfgraph*}(43,20)
\fmfleft{t1,v1,v3,t2} \fmfright{t3,v2,v4,t4} \fmfset{dot_len}{1mm} \fmf{plain,tension=1.5}{v1,v2}     \fmf{dots,tension=1.5}{v3,v4}   
 \end{fmfgraph*}}
\end{cases}, \label{eq:correlations}
\end{equation} 
Both the numerators for $C_{\pm}$ and the denominators for $G_{\pm}$, $C_{\pm}$ have corrections in higher powers of $|\mathbf{q}|^2$, 
starting with $|\mathbf{q}|^4$. In the disordered phase $\bar{\rho}<\bar{\rho}^*$, the quartic momentum terms are irrelevant. 
We will be interested in finding fluctuation corrections that generate a peak in the structure factor, along with 
a possible kink near the origin, which cannot exist in a mean-field theory.  The reason is simple: 
Any odd terms in $q_{\parallel}$ are always imaginary and give contributions to the velocity terms proportional to the drive $\delta$. 
Therefore, they are always removed by an appropriate choice of co-moving frame for the fields $\phi_{\pm}$. We can see this explicitly when 
we calculate the static structure factors $S_{\rho,\psi}(\mathbf{q})$ directly from the correlation functions $C_{\pm}$.
\begin{eqnarray}
S_{\rho,\psi}(\mathbf{q}) &=&\int_{-\infty}^{\infty} \frac{\mathrm{d}\omega}{2\pi}  \frac{2f^{(\pm)}_N(\mathbf{q}) }{\left[\omega-f^{(\pm)}_v(\mathbf{q}) \right]^2+\left[f^{(\pm)}_D(\mathbf{q})\right]^2} \nonumber\\
&= &\frac{f^{(\pm)}_N(\mathbf{q}) }{|f^{(\pm)}_D(\mathbf{q})|} \label{eq:Sqintegral}
\end{eqnarray}
In this  expression, we have included the higher-order corrections for $C_{\pm}$ appearing in the velocity 
[$f^{(\pm)}_v(\mathbf{q}) \equiv v_{\pm} q_{\parallel}+C^{(\pm)}_v|\mathbf{q}|^2q_{\parallel}+\ldots $], the noise 
[$f^{(\pm)}_N(\mathbf{q}) \equiv N_{\pm}   \bar{\rho}  |\mathbf{q}|^2+C^{(\pm)}_N \bar{\rho}|\mathbf{q}|^4+\ldots $], 
and the diffusion 
[$f^{(\pm)}_D(\mathbf{q}) \equiv D_{\pm}|\mathbf{q}|^2+C^{(\pm)}_D|\mathbf{q}|^4+C^{\times}_Dq_{\parallel}^2q_{\perp}^2+\ldots $]. Here,
$C^{(\pm)}_D \approx 3/16$ and $C_D^{\times} \approx 3/8$, to leading order in $\bar{\rho}$.
We see that $f^{(\pm)}_v(\mathbf{q})$ does not contribute to the static structure factors, so that odd powers of $\mathbf{q}$
never appear. Terms like $|q_{\parallel}|^3$ in Eq.~\eqref{eq:skphenom} must be sought from the fluctuation corrections 
induced by the interaction terms (cubic and higher-order in the fields) in the action $\mathcal{J}$ [Eq.~\eqref{eq:actionDRWLG}].

\begin{figure}[htp]
\centering
\includegraphics[width=0.36\textwidth]{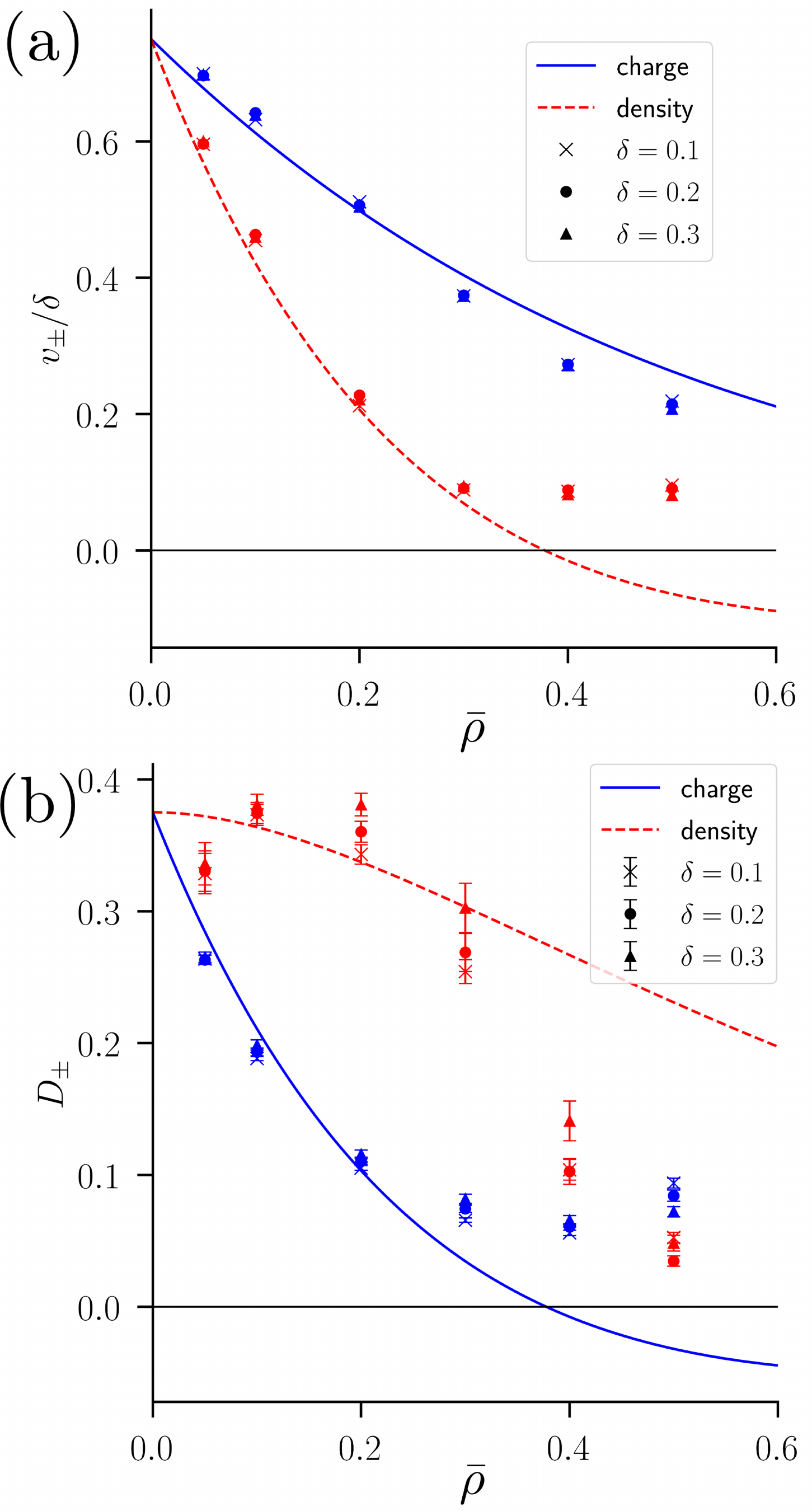}
\caption{\label{fig:velsdiffs} Velocities $v_{\pm}$ (a) and diffusivities $D_{\pm}$ (b) associated with the dynamic structure factors for the charge ($-$) and density ($+$) fields, obtained by fitting the forms in Eq.~\eqref{eq:correlations} to simulation data for two-dimensional systems ($L=50$, equal fractions of $A$ and $B$ particles) total particle density $\bar{\rho}$. The lines (red dashed and solid blue) show the theoretical results using the Doi-Peliti approach. We calculate these quantities for different values of the drive $\delta$ and show that the simulation results are consistent with the mean-field theory prediction that  $v_{\pm}$ are  proportional to $\delta$ while $D_{\pm}$   are independent of the drive $\delta$.} 
\end{figure}

Before analyzing fluctuation corrections, it is worth comparing the results of this field-theoretic approach to simulation data 
for low densities and small drive, where we expect good agreement with the ``bare'' correlation functions in Eq.~\eqref{eq:correlations}. 
Specifically, after inverting back to ($\mathbf{r},t$), they correspond to Gaussians for $S_{\psi,\rho}(\mathbf{r},t)$. In particular,
along the drive direction [$\mathbf{r}=(x_{\parallel},0)$], we find
\begin{equation}
S_{\rho,\psi}[\mathbf{r},t]=\frac{N_{\pm} \bar{\rho}}{ 4 \pi  D_{\pm}^2 t}  \,\exp\left[-\frac{(x_{\parallel}-v_{\pm} t)^2}{4D_{\pm} t}\right],
\end{equation}
with the plus (minus) signs for the density (charge) case. This form is consistent with the simulation results shown in Fig.~\ref{fig:Sxt}. 
Fitting these curves to drifting Gaussians, we can extract simulation values for $v_{\pm}/\delta$ and $D_{\pm}$. These can be compared 
with theoretical values, estimated beyond the lowest order in $\bar{\rho}$ by using the Doi-Peliti approach (see Appendix~\ref{appx:doipeliti}):
$v_{\pm}/\delta = e^{-3\bar{\rho}} [(1-\bar{\rho})(1+2 e^{\bar{\rho}/2}) \mp2\bar{\rho} (2+3 e^{\bar{\rho}/2}) ]/4$ and 
$D_{\pm} = e^{-3\bar{\rho}}(1+2e^{\bar{\rho}/2}\pm \bar{\rho}(3+5 e^{\bar{\rho}/2}))/8 $.
The comparison is shown in Fig.~\ref{fig:velsdiffs} with $v_{+}/\delta$, $D_{+}$ indicated by dashed red lines and 
$v_{-}/\delta$, $D_{-}$, by solid blue. Given that we relied on only the lowest possible order in the field-theoretic approach, 
it is remarkable how well the predictions fare,   especially at small $\bar{\rho}$. Note that at larger densities $\bar{\rho}$ that approach the phase transition, our mean-field results predict a vanishing $D_-$, which is not observed in the simulation results [see Fig.~\ref{fig:velsdiffs}(b)]. We thus expect the perturbation theory for the fluctuation corrections developed in the next section to break down near the phase transition, where other techniques (e.g., the renormalization group) have to be employed. 

We end this section with   some comments on the aspects of our driven lattice gas which may be understood in terms of a ``bare'' theory. 
The velocities of the two fields are different, with that of the density-field being lower: $v_+<v_-$. Though the theoretical value
seems to vanish at some special $\bar{\rho}$, it is unclear if this point is of any physical significance. We should remind ourselves 
that this is the velocity of the \textit{fluctuations} and not the density itself. Such is a common experience in traffic, where local jams
(fluctuations of higher density) are often observed to travel ``backwards'' from the direction of the drive. 
The two diffusion coefficients $D_{\pm}$ also differ in general. Though $D_+>0$ for all values of $\bar{\rho}$, $D_-$ vanishes at some point, as mentioned above. 
Ordinarily (in both equilibrium systems and DDS), this is a signal of criticality and onset of phase separation. However, a vanishing 
coefficient of $q^2$ cannot describe the transition into stripes observed. Instead, there must be a divergence of $S_{\rho}(\mathbf{q})$ at
$|q_{\parallel}| = 2q_{\parallel}^*$ (to accompany the divergence of $S_{\psi}$ at $|q_{\parallel}| = q_{\parallel}^*$). Nevertheless, 
for densities far below criticality, this theory does capture the drifting and diffusive behavior of both kinds of fluctuations.

Note that the difference in velocities $v_d=v_+-v_-$ being non-zero is a key feature of the model. As a result, it is not possible to 
choose a co-moving frame in which \textit{both} densities suffer no  drift. In technical terms, we cannot eliminate both drift terms
from the Langevin equations for $\phi_{\pm}$, Eq.~\eqref{eq:langevin}, so that one of them must be involved when we consider 
the fluctuation corrections. The presence of a drift term is not only a major difference between the field-theoretic formulation for our model 
and that for DDS \cite{KLSFT1,KLSFT2}, but also appears to be the key ingredient for the emergence of spatial structures periodic along the drive.

To summarize, mean-field theory (i.e., the quadratic part of $\mathcal{J}$) provides us with static structure factors of the form
\begin{equation}
S_{\rho,\psi}(\mathbf{q})= \frac{N_{\pm} \bar{\rho}|\mathbf{q}|^2}{D_{\pm}|\mathbf{q}|^2+C^{(\pm)}_D |\mathbf{q}|^4+C^{\times}_Dq_{\parallel}^2 q_{\perp}^2}, \label{eq:Sqmeanfield}
\end{equation}
which is clearly inadequate for generating the characteristic ``microemulsion'' peaks in $S_{\psi}$ when $\delta>0$. These peaks can only 
arise from the fluctuation corrections (with $v_d \ne 0$ playing a key role), which will be our focus in the next section.

\section{Fluctuation corrections \label{sec:coarsen}}

Corrections due to fluctuations can be developed using a perturbative
approach, starting with the   quadratic terms in the action $\mathcal{J}$
[i.e., a ``free'' field theory, with the
correlation functions in Eq.~\eqref{eq:correlations}]. In this work, we
restrict our attention to how these two-point functions are modified by the
cubic terms in $\mathcal{J}$, i.e., those with $g_{0,\pm}\propto\delta$:
$\parbox{1.3cm}{\begin{fmfgraph*}(40,20)
\fmfright{v1} \fmfleft{v2,v3} \fmf{plain,tension=1}{w1,v2} \fmfset{arrow_len}{2mm} \fmfset{dot_len}{1mm} \fmfset{arrow_ang}{20} \fmf{dots,tension=1.5}{w11,w1} \fmf{plain,tension=1}{w1,v3} \fmf{phantom_arrow,tension=2}{w11,v1}    \end{fmfgraph*}} $ for $g_-$,   $\parbox{1.3cm}{\begin{fmfgraph*}(40,20)
\fmfright{v1} \fmfleft{v2,v3} \fmf{plain,tension=1}{w1,v2} \fmfset{arrow_len}{2mm} \fmfset{dot_len}{1mm} \fmfset{arrow_ang}{20} \fmf{plain,tension=1.5}{w1,w11} \fmf{dots,tension=1}{w1,v3} \fmf{phantom_arrow,tension=2}{w11,v1}   \end{fmfgraph*}}$  for $g_0$, and   $\parbox{1.3cm}{\begin{fmfgraph*}(40,20)
\fmfright{v1} \fmfleft{v2,v3} \fmfset{arrow_len}{2mm} \fmfset{arrow_ang}{20} \fmfset{dot_len}{1mm} \fmf{dots,tension=1}{w1,v2} \fmf{dots,tension=1.5}{w1,w11} \fmf{dots,tension=1}{w1,v3} \fmf{phantom_arrow,tension=2}{w11,v1} \end{fmfgraph*}}$  for $g_+$, with the arrows indicating the response fields on which we place the derivative $\partial_{\parallel}$.  Further, we
will   focus on finding corrections to just the \textit{static} structure
factors for $d=2$. Thus, we will concentrate only on the self-energies
$\Sigma_{\pm}\equiv\Sigma_{\pm}(\mathbf{q},\omega)$ (which will correct the
propagators for the density and charge fields, respectively) and the
corresponding corrections to the noise correlations, $\eta_{\pm}\equiv
\eta_{\pm}(\mathbf{q},\omega)$. These enter into the fluctuation-corrected
causal propagators $\bar{G}_{\pm} \equiv \bar{G}_{\pm}(\mathbf{q},\omega)$ via Dyson's equation:
\begin{align}
\bar{G}_{\pm} ^{-1}  &  =G_{\pm}%
^{-1}-\Sigma_{\pm}\nonumber\\
&  =-i(\omega-v_{\pm}q_{\parallel}+\Im\Sigma_{\pm})+D_{\pm}|\mathbf{q}%
|^{2}-\Re\Sigma_{\pm} \label{eq:Dyson}%
\end{align}
and the corrected noise correlations via%
\[
\bar{N}_{\pm}(\mathbf{q},\omega)=N_{\pm}|\mathbf{q}|^{2}+\eta_{\pm}%
(\mathbf{q},\omega).
\]
These corrections contribute to the static structure factors through the expression:%
\begin{align}
&  S_{\rho,\psi}(\mathbf{q})=\int\frac{\mathrm{d}\omega}{\pi}%
\Bigg[\,\nonumber\\
&  \frac{N_{\pm}|\mathbf{q}|^{2}+\eta_{\pm}}{(\omega-v_{\pm}q_{\parallel}%
+\Im\Sigma_{\pm})^{2}+(D_{\pm}|\mathbf{q}|^{2}-\Re\Sigma_{\pm})^{2}}\Bigg].
\label{eq:Sqintegral2}%
\end{align}
Note that if we ignore the $\omega$ dependence in $\Sigma$ and $\eta$, then
the integration over $\omega$ eliminates $v_{\pm}q_{\parallel}-\Im\Sigma_{\pm
}$ and has the same effect as evaluating the corrections in the co-moving
frame $\omega=v_{\pm}q_{\parallel}$ while keeping only $\Re\Sigma_{\pm}$. In
the analysis presented below, we will avoid such an uncontrollable
approximation and compute the integral numerically instead. The details of
this calculation are quite involved and are deferred to Appendix
\ref{appx:loopdetails} for the interested reader. In this section, we will
provide a brief overview of the various ingredients and steps, ending with the
results and a discussion of their implications.

In two dimensions, we can generally expect (logarithmic) divergence at both
the UV and IR ends. As we plan to compare our results to simulation data on a
finite lattice, we will simply impose an UV cutoff, $\Lambda$, and evaluate at
non-zero wavenumbers [specifically, $q_{\parallel}\sim O(1/L)$]. In fact, since
both densities are conserved, $S_{\rho,\psi}(\mathbf{q}=0)$ are fixed \cite{DRWLG1}.

Turning first to the self-energy $\Sigma_{-}\equiv\ \Sigma_{-}(\mathbf{q}%
,\omega)$ for the charge field, we see that the lowest order corrections (i.e.,
with one loop, $\propto$ $\delta^{2}$) are, in terms of diagrams, given by
\begin{align}
& \Sigma_-(\mathbf{q},\omega) = \parbox{2.1cm}{ \begin{fmfgraph*}(60,50) \fmfleft{v1} \fmfright{v2} \fmfset{dot_len}{1mm} \fmfset{arrow_len}{2mm} \fmf{plain_arrow,tension=3}{v1,w1} \fmf{dots_arrow,left,tension=1}{w1,w2} \fmf{plain,right,tension=1}{w1,w2} \fmf{plain_arrow,tension=3}{w2,v2}  \end{fmfgraph*}  }+\parbox{2.1cm}{ \begin{fmfgraph*}(60,50) \fmfleft{v1} \fmfright{v2} \fmfset{dot_len}{1mm} \fmfset{arrow_len}{2mm} \fmf{plain_arrow,tension=3}{v1,w1} \fmf{plain_arrow,left,tension=1}{w1,w2} \fmf{dots,right,tension=1}{w1,w2} \fmf{plain_arrow,tension=3}{w2,v2}  \end{fmfgraph*}  }   \label{eqn:SE-diagrams}\\
&  =\int\mathrm{d}\omega_{k}\,\mathrm{d}\mathbf{k}\,\left\{  g_{0}g_{-}%
...+g_{0}^{2}...\right\}   \label{eq:loopintegral0}.
\end{align}
  The key feature of the loop integration is the mixing of the propagators for
the charge and density fields [dashed and solid lines in
Eq.~\eqref{eqn:SE-diagrams}] which prevents us from eliminating the linear
drift terms $v_{\pm}k_{\parallel}$ in both propagators simultaneously. In the
usual single-species DDS case, for example, a Galilean transformation  into a co-moving frame removes the linear drift term. This is not
possible here as the charge and density fields have \textit{different}
characteristic velocities $v_{\pm}$. Indeed, as will be shown below,
$v_{d}\equiv v_{+}-v_{-}$ plays the key role for characterizing the periodic
structures. A similar phenomenon occurs in a model for drifting crystals
treated in Ref.~\cite{driftingcrystal1}, which also has two scalar fields with
different drift velocities.

Substituting in the propagators from Eq.~\eqref{eq:correlations} and evaluating the
integrals, we find, to leading order in the average density $\bar{\rho}$,
\begin{align}
\Sigma_{-}  &  \approx g_{0}\frac{\bar{\rho}\bar{N}v_{d}q_{\parallel}}%
{16\pi\bar{D}^{3}}\Bigg\{\left[  \frac{2(g_{-}-g_{0})q_{\parallel}}{q_{c}%
}-\frac{i(g_{-}+g_{0})}{2}\right] \nonumber\\
&  \times\ln\left[  \frac{\left(  4\Lambda\right)  ^{2} }{2i(\bar
{v}q_{\parallel}-\omega)/\bar{D}+q_{c}^{2}+|\mathbf{q}|^{2}}\right]
+i(g_{-}+g_{0})\Bigg\}, \label{eq:loopintegral} 
\end{align}
where 
\begin{equation}
q_{c}\equiv\frac{v_{d}}{2\bar{D}} \label{qcross} 
\end{equation}
is a crossover wavenumber that will be prominent in our discussions below.
From this complex expression, it is instructive to study the real and
imaginary parts of $\Sigma_{-}$ (with real $\mathbf{q}$ and $\omega$)
separately, as they enter into the diffusive and drift parts of
Eq.~(\ref{eq:Sqintegral2}), respectively. As discussed above, $\Re\Sigma_{-}$
plays a more significant role for $S_{\psi}$ and it reads%
\begin{align}
&  \Re\Sigma_{-}=\frac{\bar{N}\bar{\rho}g_{0}}{32\pi\bar{D}^{2}}%
\Bigg\{\nonumber\\
&  ~(g_{-}-g_{0})q_{\parallel}^{2}\ln\left[  \frac{\left(  4\Lambda\right)
^{4}}{[2(\bar{v}q_{\parallel}-\omega)/\bar{D}]^{2}+\left(  q_{c}%
^{2}+|\mathbf{q}|^{2}\right)  ^{2}}\right] \nonumber\\
&  ~~~-2(g_{-}+g_{0})q_{c}q_{\parallel}\tan^{-1}\left[  \frac{2(\bar
{v}q_{\parallel}-\omega)/\bar{D}}{q_{c}^{2}+|\mathbf{q}|^{2}}\right]  \Bigg\}.
\end{align}
The first term in the curly braces, proportional to $q_{\parallel}^{2}$, would
renormalize the diffusivity $D_{-}$ in the drive direction. It introduces a
DDS-like anisotropy and leads to a potential discontinuity in the static
structure factor at $\mathbf{q}=0$. The second term, with coefficient
$q_{c}q_{\parallel}\propto v_{d}q_{\parallel}$, has no DDS analog. Of course,
an integration over $\omega$ is needed to obtain the static $S_{\psi}\left(
\mathbf{q}\right)  $. But, as noted above, we can find a rough estimate of its
effect by evaluating this term in the co-moving frame (by setting
$\omega=v_{-}q_{\parallel}$ in this case). The result is $\propto
q_{c}q_{\parallel}\tan^{-1}\left[  q_{c}q_{\parallel}/\left(  q_{c}%
^{2}+|\mathbf{q}|^{2}\right)  \right]  $, confirming that it is necessarily
even in $q_{\parallel}$. Yet, it provides the signal of a crossover: For small
$q_{\parallel}$, it starts as $q_{\parallel}^{2}$, but it behaves as
$|q_{\parallel}|$ for values around $q_{c}$ (where the $\tan^{-1}$ is slowly
varying). This is the property that offers the possibility of a maximum in
$S_{\psi}$, as we shall see. By contrast, the imaginary part, $\Im\Sigma_{-}$,
does not affect $S_{\psi}$ qualitatively and the result may be found
Appendix~\ref{appx:loopdetails}.

Turning to the correction to the noise in $\bar{N}_{-}(\mathbf{q}%
,\omega)=N_{-}\bar{\rho}|\mathbf{q}|^{2}+\eta_{-}$, we see that the relevant
diagram is 
\begin{equation}
\eta_{-} 
=\parbox{2.5cm}{ \begin{fmfgraph*}(70,50) \fmfleft{v1} \fmfright{v2} \fmfset{dot_len}{1mm} \fmfset{arrow_len}{2mm} \fmf{plain_arrow,tension=2}{w1,v1} \fmf{dots,left,tension=1}{w1,w2} \fmf{plain,right,tension=1}{w1,w2} \fmf{plain_arrow,tension=2}{w2,v2}  \end{fmfgraph*}  }=\int%
\mathrm{d}\omega_{k} \,\mathrm{d}\mathbf{k}\,\left\{  g_{0}^{2}...\right\}
\label{eqn:N-diagrams} 
\end{equation}
with the result 
\begin{equation}
\eta_{-}\approx g_{0}^{2}\frac{\bar{N}^{2}\bar{\rho}^{2}}{8\pi\bar{D}^{3}%
}q_{\parallel}^{2}\ln\left[  \frac{\left(  4\Lambda\right)  ^{4}}{[2(\bar
{v}q_{\parallel}-\omega)/\bar{D}]^{2}+\left(  q_{c}^{2}+|\mathbf{q}%
|^{2}\right)  ^{2}}\right]  .
\end{equation}
As may be expected, $\eta_{-}$ is real and, being proportional to
$q_{\parallel}^{2}$, introduces an anisotropy in the noise correlations. Note
that every correction vanishes with $q_{\parallel}$ so that $S_{\psi} (
q_{\parallel}=0,q_{\bot} )  $ suffers no modifications (at least at this
lowest order). In other words, there are no changes for the structure factors
perpendicular to the drive. This is a feature our model shares with the
single-species DDS, leading to a discontinuity singularity at $ 
\mathbf{q}=0 $.

  \begin{figure}[htp]
\centering
\includegraphics[width=0.4\textwidth]{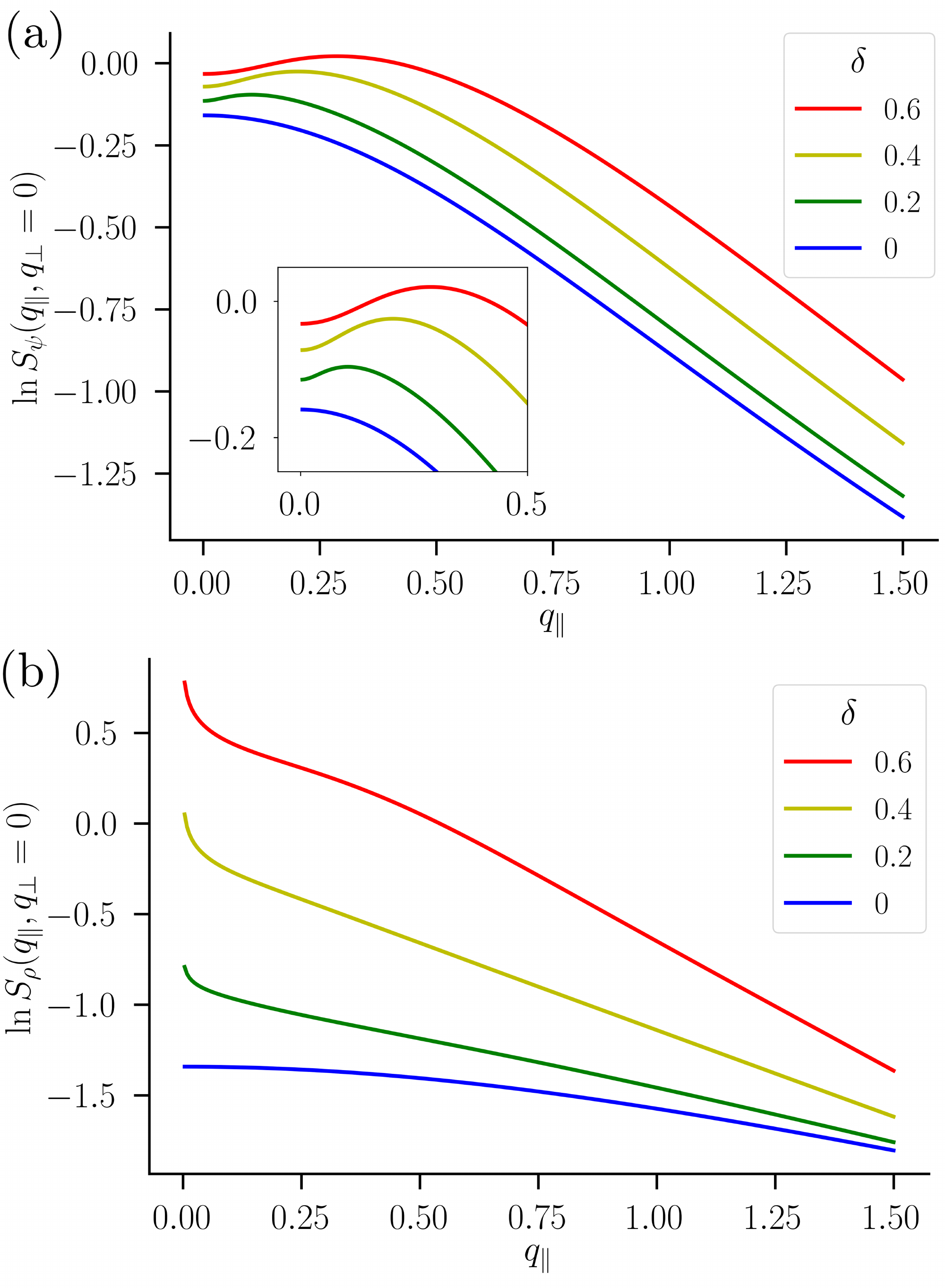}
\caption{\label{fig:stheory} Static  structure factors for (a) the charge fields $S_{\psi}(\mathbf{q})$ and (b) the particle density $S_{\rho}(\mathbf{q})$ [evaluated in two dimensions along the drive direction $\mathbf{q}=(q_{\parallel},q_{\perp}=0)$] for different values of the drive $\delta$ and a fixed $\bar{\rho}=0.2$, as calculated from the one loop corrections to  the propagator and the noise correlations.  Note the qualitative agreement between these structure factor forms and the ones found in simulations in Fig.~\ref{fig:sfactor}(a) and Fig.~\ref{fig:densfactor}.} 
\end{figure}

When all these corrections are inserted into Eq.~(\ref{eq:Sqintegral2}) and
the $\omega$ integration performed numerically, we find that the modification
to the structure factor is positive, as illustrated by the $\delta\neq0$
curves for $S_{\psi} (  q_{\parallel},q_{\bot}=0 )  $ in
Fig.~\ref{fig:stheory}(a). Thus, we recover the DDS-like discontinuity
observed in simulations (see Fig.~\ref{fig:Skfull}): $S_{\psi} (
q_{\parallel}\rightarrow 0,q_{\bot}=0 )  >S_{\psi} (  q_{\parallel
}=0,q_{\bot}\rightarrow0 )  $. More crucially, as we examine $S_{\psi
} (  q_{\parallel},q_{\bot}=0 )  $\ at larger $q_{\parallel}$, we
find that the effectively $|q_{\parallel}|$ behavior in the fluctuation
corrections creates a peak at a non-trivial $q_{\parallel}^{\ast}>0$.
However, the theoretical result does not have a $|q_{\parallel}|$-like kink at
the origin, but crosses over to a smooth $q_{\parallel}^{2}$. This
``rounding off'' is especially evident for
the $\delta=0.6$ case [inset of Fig.~\ref{fig:stheory}(a)].  By contrast, there
is no hint of such smoothing in the data, Fig.~\ref{fig:sfactor}(a) (although
this feature may reveal itself in larger systems). Turning to the peak
position, our theory finds $q_{\parallel}^{\ast}\approx0.398|q_{c}%
|\approx1.26\bar{\rho}|\delta|$. Note that it is proportional to the drive
magnitude $|\delta|$ and the particle density $\bar{\rho}$. To compare this
expression with data, we obtain a linear fit to the red points in
Fig.~\ref{fig:stripesize} (2D) for small $|\delta|$ and found $q_{\parallel
}^{\ast}\approx0.64|\delta|$ (red line). Since the data are from a system with
$\bar{\rho}=0.5$, the theoretical value would be $q_{\parallel}^{\ast}%
\approx0.63|\delta|$ , which is in surprisingly good agreement with simulation
results. Of course, this good agreement may be coincidental, as only the
lowest orders in $\bar{\rho}$ and $\delta$ have been kept in the field
theoretic treatment.

Finally, we consider the corrections to the static structure factors for the
density fields, $S_{\rho}(\mathbf{q})$. Again, we defer the details of the
calculation to Appendix~\ref{appx:loopdetails} and present only some
highlights here. The graphs for the self-energy and noise corrections are
similar to those above:
\begin{align}
\Sigma_{+} &
=\parbox{2.2cm}{ \begin{fmfgraph*}(60,50) \fmfleft{v1} \fmfright{v2}  \fmfset{dot_len}{1mm} \fmfset{arrow_len}{2mm} \fmf{dots_arrow,tension=1.5}{v1,w1} \fmf{plain_arrow,left,tension=0.7}{w1,w2} \fmf{plain,right,tension=0.7}{w1,w2} \fmf{dots_arrow,tension=1.5}{w2,v2} \end{fmfgraph*} }+\,\parbox{2cm}{ \begin{fmfgraph*}(60,50) \fmfleft{v1} \fmfright{v2} \fmfset{arrow_len}{2mm} \fmfset{dot_len}{1mm} \fmf{dots_arrow,tension=1.5}{v1,w1} \fmf{dots_arrow,left,tension=0.7}{w1,w2} \fmf{dots,right,tension=0.7}{w1,w2} \fmf{dots_arrow,tension=1.5}{w2,v2} \end{fmfgraph*} }\label{eq:Sigma+pix}
\\[-15pt] \eta_{+} &
=\parbox{2.2cm}{ \begin{fmfgraph*}(60,50) \fmfleft{v1} \fmfright{v2} \fmfset{arrow_len}{2.3mm} \fmfset{dot_len}{1mm} \fmf{dots_arrow,tension=1.5}{w1,v1} \fmf{dots,left,tension=0.7}{w1,w2} \fmf{dots,right,tension=0.7}{w1,w2} \fmf{dots_arrow,tension=1.5}{w2,v2}  \end{fmfgraph*}  }+\,\parbox{2cm}{ \begin{fmfgraph*}(60,50) \fmfleft{v1} \fmfright{v2} \fmfset{dot_len}{1mm} \fmfset{arrow_len}{2.3mm}\fmf{dots_arrow,tension=1.5}{w1,v1} \fmf{plain,left,tension=0.7}{w1,w2} \fmf{plain,right,tension=0.7}{w1,w2} \fmf{dots_arrow,tension=1.5}{w2,v2}  \end{fmfgraph*}  }\label{eq:eta+pix}%
\end{align}
and provide us with
\begin{align*}
\Sigma_{+} &  =\int\mathrm{d}\omega_{k}\,\mathrm{d}\mathbf{k}\,\left\{
g_{0}g_{-}...+g_{+}^{2}...\right\}  \\
\eta_{+} &  =\int\mathrm{d}\omega_{k}\,\mathrm{d}\mathbf{k}\,\left\{  g_{+}
^{2}...+g_{-}^{2}...\right\}.
\end{align*}
The results are 
\begin{align}
\Sigma_{+} &  =\frac{\bar{\rho}q_{\parallel}^{2}}{16\pi}\Bigg\{g_{0}g_{-}
\frac{N_{-}}{D_{-}^{2}}\ln\left[  \frac{\left(  4\Lambda\right)^{2}  
e^{-i\pi/2}}{2(v_{-}q_{\parallel}-\omega)/D_{-}-i|\mathbf{q}|^{2}}\right]
\nonumber\\
&  {}-g_{+}^{2}\frac{N_{+}}{D_{+}^{2}}\ln\left[  \frac{\left(  4\Lambda
\right)^{2}}{2i(v_{+}q_{\parallel}-\omega)/D_{+}+|\mathbf{q} 
|^{2}}\right]  \Bigg\}\label{eqn:Sigma+} 
\end{align}
and  \begin{align}
\eta_{+} &  =\frac{\bar{\rho}^{2}q_{\parallel}^{2}\,}{32\pi}\sum_{\alpha=\pm
}\Bigg\{\frac{g_{\alpha}^{2}N_{\alpha}^{2}}{D_{\alpha}^{3}}\nonumber\\
&  \quad\times\ln\left[  \frac{\left(  4\Lambda\right)  ^{4}}{\left[
2(v_{\alpha}q_{\parallel}-\omega)/D_{\alpha}\right]  ^{2}+|\mathbf{q}|^{4}}\right]
\Bigg\}.\label{eq:eta+} 
\end{align}

 \begin{figure}[htp]
\centering
\includegraphics[width=0.435\textwidth]{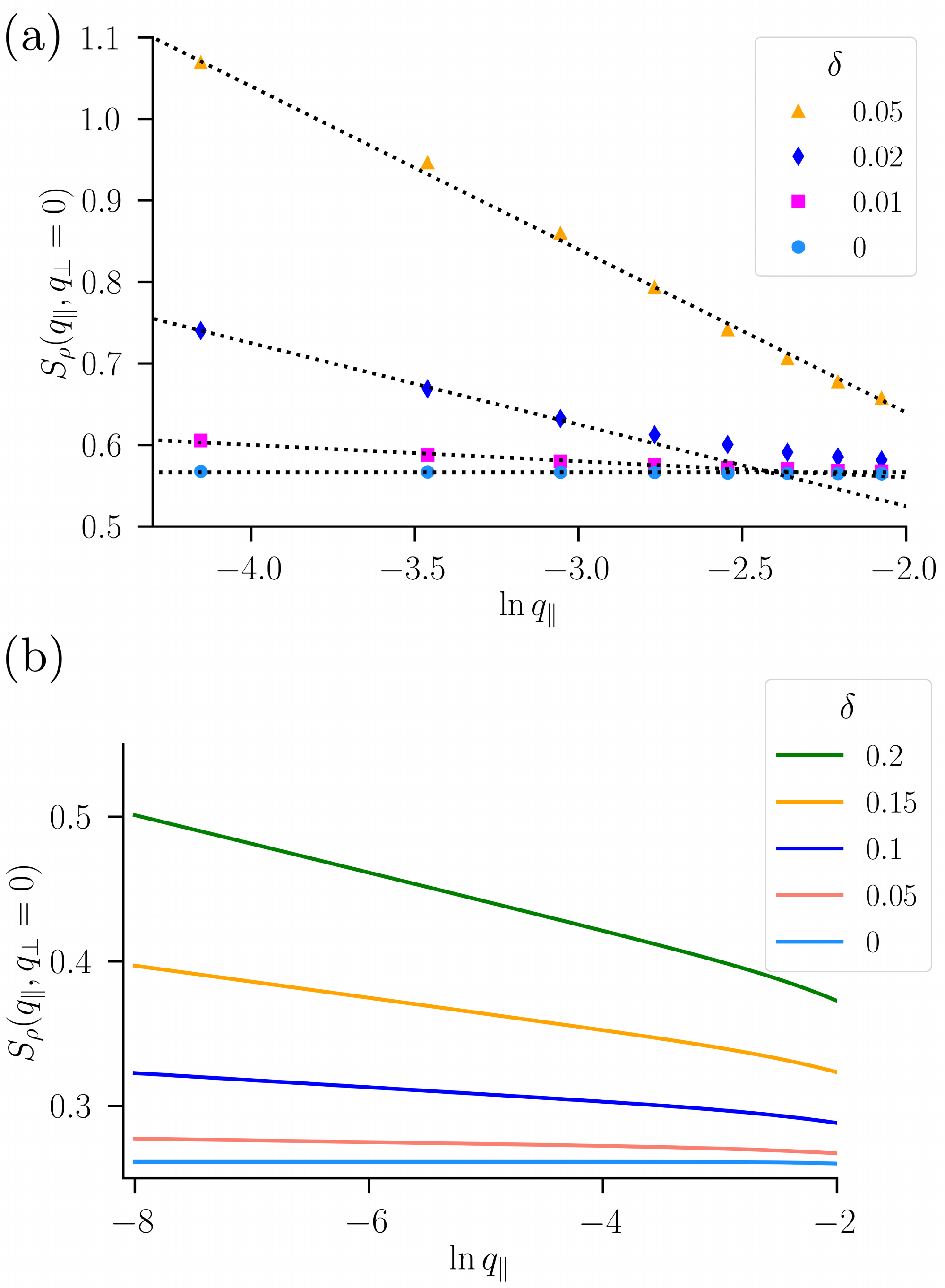}
\caption{\label{fig:logDiv} { Static structure factors $S_{\rho}(\mathbf{q})$ at small $q_{\parallel}$ for simulation data in (a) [from systems with $L^2=400^2$ lattice sites at an average particle density of $\bar{\rho}=0.5$, as in Fig.~\ref{fig:densfactor}] and the field-theoretic results in (b) [obtained via numerical integration of Eq.~\eqref{eq:Sqintegral2} for $\bar{\rho}=0.2$, as in Fig.~\ref{fig:stheory}(b)]. Note the qualitative agreement between these results: The linear behavior of $S_{\rho}$ with $\ln q_{\parallel}$ at small $\ln q_{\parallel}$ values for non-zero drive $\delta>0$ in both (a) and (b) is indicative of a logarithmic divergence, predicted by the field theory, which vanishes when $\delta=0$, where we find a horizontal line (light blue lines and points). The straight dashed black lines in (a) are a guide to the eye to illustrate the linear behavior for small $\ln q_{\parallel}$.    }  } 
\end{figure}

It is instructive to compare these corrections to the ones for the charge fields, $\Sigma_{-}$\ and $\eta_{-}%
$. In the latter, the small $\omega$, $\mathbf{q}$ behavior is regulated by
$q_{c}$ (i.e., the velocity difference $v_{d}$). By contrast, we find
logarithmic divergences in the IR regime for $\Sigma_{+}$\ and $\eta_{+}$.
When inserted in Eq.~(\ref{eq:Sqintegral2}) and the integration carried out,
these generate the cusp-like structures in $S_{\rho}$ as $q_{\parallel
}\rightarrow0$. In Fig.~\ref{fig:stheory}(b), we illustrate $S_{\rho} (
q_{\parallel},0 )  $ for various drives. Quantitatively, there is little
agreement between these $S_{\rho}$'s and the simulation values in
Fig.~\ref{fig:sfactor}(b). Nevertheless, we call attention to two important
qualitative features: One is that, when driven, the fluctuations modify the
small $q_{\parallel}$ behavior considerably. Instead of a kink with a positive slope
[downwards as $q_{\parallel}\rightarrow0$, in Fig.~\ref{fig:sfactor}(a)] for the
$S_{\psi}$'s, the data here show a \textit{cusp}\footnote{We have reasons (see
below) to believe that $S_{\rho}$ diverges as $\ln q$, so that the slope will
diverge as $q\rightarrow0$. Thus, we use the term ``cusp'' instead of ``kink'' here.
 } with a large negative slope (upwards), which is especially prominent in
the $\delta=0.02$ case. This feature is certainly displayed in the theoretical
$S_{\rho}$'s. Indeed, it is straightforward to uncover an IR divergence
associated with the integrals regularized by $q_{\parallel}\neq0$. In this case, we
expect a singularity of the form of $\ln q_{\parallel}$ in  $S_{\rho}$.  To probe this property further, we plot $S_{\rho
} (  q_{\parallel},0 )  $ \textit{vs.} $\ln q_{\parallel}$ , as illustrated in Fig. \ref{fig:logDiv},
 for both the data (a) and theoretical results (b). Surprisingly, for
small $q_{\parallel}$, both display a behavior linear in $\ln q_{\parallel}$! We caution
that such a behavior should not be extrapolated naively. It is
likely that more subtle physics comes into play at larger scales \textit{and}
that higher order corrections will be important.

Note that these divergences ($\ln q_{\parallel}$ in $d=2$) are reminiscent of the
behavior of the single species driven diffusive system in the disordered
phase. There, the static structure factors are regular in the IR limit, while
the singularities appears in dynamics, as anomalous diffusion for $d\leq
d_{c}=2$ \cite{KLSFT3,AnomalousDiff}. We believe it is the presence of two
different drift velocities in our model (which prevents us from studying the system in a co-moving frame) which allows the dynamic singularities to induce
the ones in static quantities. It would be interesting, but beyond the scope
of this paper, to find such connection and to explore how anomalous diffusion
manifests in the DWRLG.

Before ending this section, note that the $S_{\rho}$'s [in
Fig.~\ref{fig:densfactor}  or \ref{fig:stheory}(b)] do not display peaks like
the ones in the charge $S_{\psi}$'s. Yet, as the overall density approaches the
critical value and stripes form, $S_{\psi}(  q_{\parallel}^{\ast},0)  $
will diverge \textit{and} so must $S_{\rho}(  2q_{\parallel}^{\ast},0) $.
Thus, peaks in the latter must develop. We believe that, for the values of
$\bar{\rho}$ and $\delta$ shown here, such peaks are shrouded by the
logarithmic divergence near the origin. Indeed, for the largest drives
illustrated, there is a detectable shoulder in the theoretical
$S_{\rho}$ [red curve for $\delta=0.6$ in Fig.~\ref{fig:stheory}(b)] and the hint of one in the data [purple curve for $\delta=0.5$ in
Fig.~\ref{fig:densfactor}]. We see that the leading order corrections
considered here reproduce the qualitative features of the structure factors
for both the charge and density fields. In this sense, we believe that the
approach developed here is a sound first step towards a quantitatively
successful theory. The next steps would take into account the renormalization
of all of the relevant couplings in Eq.~\eqref{eq:actionDRWLG}. Although beyond
the scope of the current work, pursuing a systematic analysis is a worthy goal.
When completed, we are confident that the surprising properties of the DWRLG
can be understood.

\section{Summary and Outlook \label{sec:conclusion}}

    Exploiting Monte Carlo and field-theoretic techniques, we analyzed a
strongly-driven lattice gas with two species ($A,B$) -- the Widom Rowlinson
model. Restricting ourselves to systems with equal numbers of each species,
there are only two control parameters: the overall particle density,
$\bar{\rho}$, and the drive, $\delta$. As the system settles into
non-equilibrium steady states, many unexpected properties emerge. As both control parameters
increase, the homogeneous phase gives way to phase-separation, with order
stripes (slabs in 3D) \textit{perpendicular} to the drive. The precursor of
this transition appears as a ``patterned'' disordered phase, characterized by a non-trivial wavelength associated with
the drive direction. Specifically, the static structure factor $S_{\psi
}(  q_{\parallel},0) $ (for the difference in local densities:
$\psi=\rho_{A}-\rho_{B}$) has a peak at $q_{\parallel}^{\ast}>0$. As
$q_{\parallel}\rightarrow0$  and we move away from the peak, $S_{\psi}$ decreases, ending with a
kink singularity (non-zero slope) at the origin. The other structure factor
$S_{\rho}$\ (for the sum: $\rho=\rho_{A}+\rho_{B}$) displays no similar peak
(far from the transition), but \textit{increases} sharply as $q_{\parallel
}\rightarrow0$ so that the singularity at the origin is \textit{cusp}-like
(diverging as $\ln q_{\parallel}$, approximately). Following standard methods
of stochastic field theory, we derived a continuum, coarse-grained
description. At the Gaussian level, the \textit{dynamic }correlation functions
provide  good agreement with the drift velocities of fluctuations. However, the Gaussian field theory fails to produce the novel properties of the static $S_{\rho,\psi}$. Taking
into account corrections at the one-loop level (i.e., keeping the lowest
non-trivial orders in $\bar{\rho}$ and $\delta$), the field theory captures
the essentials of these features (peaks in $S_{\psi}$, cusps in $S_{\rho}$).
However, it compares poorly with data quantitatively.

The driven Widom Rowlinson lattice gas differs from the well-known driven
Ising case in one key feature. In the latter, the drive produces a single
drifting $v$ associated with the local density field. By contrast, due to the
excluded volume interactions (between $A$ and $B$), the drive in the DWRLG
induces two \textit{different} drift velocities (for $\rho,\psi$). The peak in
$S_{\psi}$ can be traced to the ratio of this velocity difference and the
average diffusivity: $v_{d}/\bar{D}$. For deeper reasons we have yet to find,
this aspect also appears to allow the dynamic singularities associated with
anomalous diffusion \cite{KLSFT3,AnomalousDiff} (logarithmic in $d=2$) to
emerge in the static $S_{\rho}$. In this sense, there are many interesting
directions to explore in the future, including a comprehensive analysis of the
renormalization of the couplings of the theory.

We are aware of many puzzling features in this system which our simple theory
cannot accommodate. The high density $\bar{\rho}$ behavior, in particular, is not accessible to the perturbative approach developed here. Interesting examples at these higher densities include    the development of a peak in $S_{\rho}$ for larger $\bar{\rho}$ and $\delta$, so that  $S_{\rho}(  2q_{\parallel}^{\ast
},0)  $ diverges at the critical $\bar{\rho}^*$ [to accompany the
divergence of $S_{\psi}(  q_{\parallel}^{\ast},0)$],  and  the merging of stripes at densities above    $\bar{\rho}^*$ \cite{DRWLG1} (i.e., decreasing $q_{\parallel}^{\ast}$). We are
studying possible ways to improve the theoretical treatment along these lines.
Clearly, it would be ideal if a full-scale renormalization group analysis can
be carried out to describe the critical behavior observed in Ref.
\cite{DRWLG1}. On the simulations front, improved studies with larger systems
are underway so that various exponents, as well as the likely presence of
anisotropic scaling, can be better measured. There is also much of interest in
system at densities larger than criticality. In the previous
study \cite{DRWLG1}, interfaces are found to display Edwards-Wilkinson-like
behavior \cite{EWModel}. How can such behavior be understood, as these interfaces are drifting
due to the drive and \textit{not} part of a system in thermal equilibrium?
Apart from understanding how (and at what densities) configurations with $N$
stripes can evolve into $N-1$ stripes, we may consider the extreme high
density limit. There, the system consists of just two regions with particles
of $A$ or $B$ only (separated by thin lines of vacancies) and a handful of
holes drifting through. These ``drifters'' interact with the interfaces, of course. Other than that, their travels
through the ``bulk'' regions must be
identical to the ordinary single-species DDS. Their interactions with the
interfaces drive the latter, in a manner reminiscent of the Eden model of
growing interfaces \cite{Eden}. Simulation studies can be readily carried out, but our
hope is that some theoretical progress is also possible in what appears to be
a ``minimal'' system.

Beyond the particular system studied here, we should explore the vast, and
likely novel, territory associated with ``non-neutral'' systems ($\bar{\rho}_{A}\neq\bar{\rho}_{B}$).
There is also the question of whether the presence of an underlying lattice
plays a crucial role. For answers, we may explore molecular dynamics
simulations in a continuum, modeling colloidal particles with repulsive
interactions suspended in a solvent and driven with an applied force. The hope
is that connections between such computational and theoretical work can be
established with physical experimental realizations. 

\qquad

\begin{acknowledgments}
We are grateful to E. M. Horsley and Hugues Chat\'e for helpful discussions. Computational support was  provided by the University of Tennessee and Oak Ridge National Laboratory's Joint Institute for Computational Sciences. M.O.L. gratefully acknowledges partial funding from the Neutron Sciences Directorate (Oak Ridge National Laboratory), sponsored by the U.S. Department of Energy, Office of Basic Energy Sciences. 
R.D. acknowledges support from CNPq, Brazil, under grant No. 303766/2016-6.
\end{acknowledgments}

\appendix

\section{Doi-Peliti formalism \label{appx:doipeliti}}

 The probability distribution $P(\{\sigma_{\mathbf{x}}\},t)$ of observing particle configuration $\{ \sigma_{\mathbf{x}}\}$ at time $t$ is  conveniently represented by introducing  a vector $\ket{P}$ defined as
\begin{equation}
\ket{ P} = \sum_{\{ \sigma_{\mathbf{x}} \}} P(\{\sigma_{\mathbf{x}}\},t) \ket{ \{ \sigma_{\mathbf{x}}\}},
\end{equation}
where $\ket{ \{ \sigma_{\mathbf{x}}\}}$ is the lattice state in the occupation number representation. Then, the master equation   may be represented via a Schr\"odinger-like equation which reads
\begin{equation}
\partial_t \ket{P}=\mathcal{-L} \ket{P}, \label{eq:doievo}
\end{equation}
where the Liouville operator $\mathcal{L}$ (or pseudo-Hamiltonian) will depend on the rates given by Eq.~\eqref{eq:latticeupdates}. To construct the operator $\mathcal{L}$, we introduce creation and annihilation operators $a^{\dagger}_{\mathbf{x}}$ and $a_{\mathbf{x}}$, respectively. These   increase or decrease by one the occupation number of the $A$ species at site $\mathbf{x}$ and satisfy the commutation relations $[a_{\mathbf{x}} ,a^{\dagger}_{\mathbf{y}}]=\delta_{\mathbf{x},\mathbf{y}}$.  We also need an equivalent set $[b_{\mathbf{x}},b^{\dagger}_{\mathbf{y}}]=\delta_{\mathbf{x},\mathbf{y}}$ for the $B$ species, both of which will commute with the $A$ species operators. By applying the creation  operators to the state $\ket{0}$ with no particles (the vacuum), we may place as many particles as we wish at any of the lattice sites $\mathbf{x}$.  This is a problem, however, as our model is constrained by the excluded volume and particle repulsion rules, which must somehow be taken into account.

To incorporate both the excluded volume and the nearest-neighbor exclusion rule, we follow  van Wijland \cite{vanwijland} and introduce delta function operators $\delta_{\hat{n}^{A,B}_{\mathbf{x}},m}$ (where $m=0,1,2,\ldots$ and $\hat{n}^{A,B}_{\mathbf{x}}=a^{\dagger}_{\mathbf{x}} a_{\mathbf{x}},b^{\dagger}_{\mathbf{x}}b_{\mathbf{x}}$) which have the particle states $\ket{\{\sigma_{\mathbf{x}}\}}$ as eigenvectors with eigenvalues $\delta_{n_{\mathbf{x}}^{A,B},m}$, where $n_{\mathbf{x}}^{A,B}$ is the number of $A$ or $B$ particles at site $\mathbf{x}$. It is now  straightforward to write down the operator $\mathcal{L}$ as the delta function operators can guarantee that any hopping transition that violates the excluded volume or nearest-neighbor exclusion rules
 will vanish. The operator reads 
 \begin{widetext}
\begin{align}
\mathcal{L}&= \frac{1}{8 \tau} \sum_{\mathbf{x},\epsilon=\pm1}   \bigg\{(1+\epsilon\delta)(a_{\mathbf{x}-\epsilon\bm{\ell}_{\parallel}}^{\dagger} -a_{\mathbf{x}}^{\dagger} )a_{\mathbf{x}-\epsilon\bm{\ell}_{\parallel}}\delta_{\hat{n}^A_{\mathbf{x}-\epsilon\bm{\ell}_{\parallel}},1}\delta_{\hat{n}_{\mathbf{x}+\epsilon\bm{\ell}_{\parallel}}^B,0}\delta_{\hat{n}_{\mathbf{x}+\bm{\ell}_{\perp}}^B,0}\delta_{\hat{n}_{\mathbf{x}-\bm{\ell}_{\perp}}^B,0}         \nonumber \\
& \qquad \qquad {} +(a_{\mathbf{x}-\epsilon\bm{\ell}_{\perp}}^{\dagger}  -a_{\mathbf{x}}^{\dagger}   )a_{\mathbf{x}-\epsilon\bm{\ell}_{\perp}}\delta_{\hat{n}^A_{\mathbf{x}-\epsilon\bm{\ell}_{\perp}},1}\delta_{\hat{n}^B_{\mathbf{x}+\epsilon\bm{\ell}_{\perp}},0}\delta_{\hat{n}^B_{\mathbf{x}+\bm{\ell}_{\parallel}},0}\delta_{\hat{n}^B_{\mathbf{x}-\bm{\ell}_{\parallel}},0}       \nonumber \\
&+  \sum_{\bar{\epsilon}=\pm 1}   \left( 1+\delta \right)(a_{\mathbf{x}-\epsilon\bm{\ell}_{\parallel}-\bar{\epsilon}\bm{\ell}_{\perp}}^{\dagger}-a_{\mathbf{x}}^{\dagger}  )a_{\mathbf{x}-\epsilon\bm{\ell}_{\parallel}-\bar{\epsilon}\bm{\ell}_{\perp}}\delta_{\hat{n}^A_{\mathbf{x}-\epsilon\bm{\ell}_{\parallel}-\bar{\epsilon}\bm{\ell}_{\perp}},1}\delta_{\hat{n}_{\mathbf{x}+\epsilon\bm{\ell}_{\parallel} }^B,0} \delta_{\hat{n}^B_{\mathbf{x}+\bar{\epsilon}\bm{\ell}_{\perp}},0}       \Bigg\}\delta_{\hat{n}^B_{\mathbf{x}},0}\delta_{\hat{n}^A_{\mathbf{x}},0}+(a,A) \leftrightarrow( b,B), \label{eq:Louisville}
\end{align}
\end{widetext}
where $\tau$ is the lattice update time step, $\bm{\ell}_{\parallel}=\ell \hat{\mathbf{x}}$ is the lattice spacing in the drive direction and $\bm{\ell}_{\perp}= \ell \hat{\mathbf{y}}$ is the lattice spacing in the perpendicular direction.
Note that we get another set of the same terms for the $B$ particle motion which can be generated by replacing $a$ and $A$ with $b$ and $B$, respectively. Note that Eq.~\eqref{eq:doievo} with the Liouville operator given in Eq.~\eqref{eq:Louisville} represents the microscopic lattice dynamics exactly. The rates $\omega_{\mathbf{x}\rightarrow \mathbf{x}+\Delta \mathbf{x}}$ in Eq.~\eqref{eq:latticeupdates} are incorporated directly in Eq.~\eqref{eq:Louisville}, where we specialize here to the square lattice for which $N_n=8$ is the number of NN\ and NNN sites. 

To obtain the densities within the Doi-Peliti formalism, one first introduces coherent states $\ket{\phi_A}$, which are eigenstates of the annihilation operators: $a_{\mathbf{x}} \ket{\phi_A}=\phi^A_{\mathbf{x}}\ket{\phi_A}$ (with a similar relation for species $B$), where $\phi^A_{\mathbf{x}}$ is  a complex eigenvalue. The time evolution  in Eq.~\eqref{eq:doievo} can then be represented as a path integral over coherent state field configurations $\phi^{A,B}_{\mathbf{x}}$ using standard techniques \cite{tauberbook}. The coherent states may be related to the particle densities $\rho_{\mathbf{x}}^{A,B}$ at lattice site $\mathbf{x}$ via a Cole-Hopf transformation   $[\phi_{\mathbf{x}}^{A,B}]^*=e^{\hat{\rho}_{\mathbf{x}}^{A,B}}$ and $\phi_{\mathbf{x}}^{A,B}=e^{-\hat{\rho}_{\mathbf{x}}^{A,B}} \rho_{\mathbf{x}}^{A,B}$, where $\hat{\rho}_{\mathbf{x}}^{A,B}$ is the conjugate or response field to the densities $\rho_{\mathbf{x}}^{A,B}$  at site $\mathbf{x}$.   One then generates an action $\mathcal{J} \equiv \mathcal{J}[\hat{\rho}^{A,B}_{\mathbf{x}},\rho_{\mathbf{x}}^{A,B}]$ that facilitates the computation of averages over stochastic trajectories of any  functionals $\mathcal{O}[\rho_{\mathbf{x}}^{A,B}]$ just as in Eq.~\eqref{eq:averaging} in the main text. If we write the local densities as $\rho^{A,B}_{\mathbf{x}}=\bar{\rho}^{A,B}+\psi_{\mathbf{x}}^{A,B}$ at each lattice site $\mathbf{x}$, with $\bar{\rho}^{A,B}$ the average contribution  (so that $\sum_{\mathbf{x}}\psi^{A,B}_{\mathbf{x}}=0$), the Doi-Peliti action reads
\[
\mathcal{J}=\int_0^{t_f} \mathrm{d} t\sum_{\mathbf{x}} \left[  \hat{\psi}_{\mathbf{x}}^{A} \partial_t \psi_{\mathbf{x}}^{A}+\hat{\psi}_{\mathbf{x}}^{B} \partial_t \psi_{\mathbf{x}}^{B}+\mathcal{H}_{\mathbf{x}} \right] , \label{eq:DPaction}
\]
where $\hat{\psi}^{A,B}_{\mathbf{x}}$ are the response fields and $\mathcal{H}_{\mathbf{x}}$ is a local pseudo-Hamiltonian density that encodes the particle exclusions and hopping rates, derived from the analogous terms in the Liouville operator $\mathcal{L}$ in  Eq.~\eqref{eq:Louisville} by taking the expectation value of $\mathcal{L}$ with respect to the coherent states.  This development is described in standard texts, e.g., Ref.~\cite{tauberbook}.

 To write down the density $\mathcal{H}_{\mathbf{x}}$ explicitly for a two-dimensional lattice, it is convenient to  introduce discrete derivative operators of any function $f(\mathbf{x})$ defined on the lattice sites $\mathbf{x}=(x,y)$: $\Delta_{i,j} f(\mathbf{x}) \equiv f(\mathbf{x}+i\bm{\ell}_{\parallel}+j\bm{\ell}_{\perp})-f(\mathbf{x}) $, where $\bm{\ell}_{\perp,\parallel}$ are the lattice spacing along and perpendicular to the drive. 
 In terms of these discrete derivative operators, we have, dropping the $\mathbf{x}$ subscripts for notational convenience (so that $\mathcal{H} \equiv \mathcal{H}_{\mathbf{x}}$ and $\psi^{A,B} \equiv \psi^{A,B}_{\mathbf{x}}$, etc.):
 \begin{align}
\mathcal{H} &=   \frac{e^{-2(\bar{\rho}^A+\psi^A)-4( \bar{\rho}^B+\psi^B)}}{8\tau}\sum_{\epsilon =\pm1} \Bigg\{   (1-e^{ -\Delta_{\epsilon ,0}\hat{\psi}^A}) \nonumber \\& \times(1-\epsilon\delta) [\bar{\rho}^A+(1+\Delta_{\epsilon ,0})\psi^A] \nonumber \\ & \qquad \qquad \qquad \times e^{-\Delta_{\epsilon ,0}\psi^A -(\Delta_{-\epsilon ,0}+\Delta_{0,-1}+\Delta_{0, 1})\psi^B }    \nonumber \\
& {} +     (1-e^{-\Delta_{0,\epsilon} \hat{\psi}^A })[\bar{\rho}^A+(1+\Delta_{0,\epsilon} )\psi^A] \nonumber \\ & \qquad \qquad \qquad \times e^{-\Delta_{0,\epsilon}\psi^A -(\Delta_{0,-\epsilon}+\Delta_{-1,0}+\Delta_{1,0})\psi^B} \nonumber \\
&+\frac{  e^{-2(\bar{\rho}^A+\psi^A)-3(\bar{\rho}^B+\psi^B)}}{8 \tau} \sum_{\epsilon_{1,2}=\pm1 }(1- e^{ -\Delta_{\epsilon_1,\epsilon_2}\hat{\psi}^A }) \nonumber \\& \times (1-\epsilon_1\epsilon_2 \delta)  [\bar{\rho}^A+(1+\Delta_{\epsilon_1,\epsilon_2} )\psi^A ]  \nonumber \\ & \qquad \qquad \qquad \times e^{ -\Delta_{\epsilon_1,\epsilon_2}\psi^A-[\Delta_{-\epsilon_1,0}+\Delta_{0,-\epsilon_2}]\psi^B  }   \nonumber \\    & {}+   A \leftrightarrow B, \label{eq:DPHamiltonian}
\end{align}
 with the last term $A \leftrightarrow B$ indicating that we need to add   all the previous terms   with the two particle types switched.  The terms in Eq.~\eqref{eq:DPHamiltonian} have a certain logic: The first term (first three  lines) corresponds to hops along the $x$-axis, while the second term (fourth and fifth lines) represents the $y$-axis hops. Finally, the double summation over $\epsilon_{1,2}$ (lines six through eight) represents the hops to the four NNN lattice sites.
A similar formulation is straightforward for the three-dimensional case, except there are sixteen possible hopping locations.

This formulation, while derived exactly from the lattice hopping rules, is inconvenient for understanding the coarse-grained features of the system. We next need to move to a continuous description and introduce the local, coarse-grained particle densities $\rho_{A,B}(\mathbf{r})$. It will be here where the method becomes approximate, necessitating careful checks against simulation results.
 We begin with the Doi-Peliti action,  given by    Eq.~\eqref{eq:DPaction}, with the discrete pseudo-Hamiltonian in Eq.~\eqref{eq:DPHamiltonian}.   We  move to a continuous coordinate $\mathbf{x} \rightarrow \mathbf{r}$ and assume the fluctuation fields $\psi^{A,B}_{\mathbf{x}}$ are slowly varying on the scale of  the lattice spacing $\ell$, so we may replace them (and the corresponding response fields) with continuous fields $\psi^{A,B} \equiv \psi^{A,B}(\mathbf{r},t)$ at any time $t$ in the evolution. Then, the discrete derivative operators $\Delta_{i,j}$  defined above can be expanded in gradients with respect to $\mathbf{r}$. The expansion up to quadratic terms in  $\ell$ is   
\begin{equation}
\Delta_{i,j} \approx i \ell \partial_{\perp}+j \ell \partial_{\parallel}+\frac{\ell^2i^2}{2} \partial_{\parallel}^2+\frac{\ell^2j^2}{2} \partial_{\perp}^2+ij \ell^2\partial_{\parallel}\partial_{\perp}, \label{eq:Taylorseries}
\end{equation}
where $i,j$ are the integers representing the square lattice locations $\mathbf{x}=\ell(i,j)$.

Eq.~\eqref{eq:Taylorseries} is substituted into Eqs.~(\ref{eq:DPaction},\ref{eq:DPHamiltonian}) and we expand in powers of the fields and the lattice spacing $\ell$. We then perform a field redefinition and introduce the charge and density field fluctuations $\phi_{\pm} \equiv \phi_{\pm}(\mathbf{r},t)$, which are defined  via   $ \phi_{\pm} =\psi^A \pm\psi^B $, with corresponding response fields $\hat{\phi}_{\pm}=\hat{\psi}^A \pm \hat{\psi}^B$.  Then, introducing the total particle density $\bar{\rho}=\bar{\rho}^A+\bar{\rho}^B$, we find that Eq.~\eqref{eq:DPaction} reduces in the continuum limit to the action given by Eq.~\eqref{eq:actionDRWLG} for the equal density case $\bar{\rho}_A=\bar{\rho}_B=\bar{\rho}/2$, with all of the coupling definitions stated in the main text.

\section{Details of fluctuation corrections \label{appx:loopdetails}}

In this appendix we give a few other details of the computation for the
corrections to both the self-energy ($\Sigma_{\pm}$) and noise ($\eta_{\pm}$) terms.
For $\Sigma_-\equiv \Sigma_{-}(\mathbf{q},\omega)$, the two integrals, shown diagrammatically
in Eq.~(\ref{eqn:SE-diagrams}), are%
\begin{align}
 & \Sigma_{-}  =\int\frac{\mathrm{d}\omega_{k}%
\mathrm{d}\mathbf{k}}{(2\pi)^{3}}\,g_{0}\left[  \frac{k_{\parallel}^{2}}%
{2}+k_{\parallel}q_{\parallel}\right] \nonumber\\
&  \Bigg\{g_{-}C_{-}\left[  \frac{\mathbf{q}}{2}-\mathbf{k},\frac{\omega}%
{2}-\omega_{k}\right]  G_{+}\left[  \frac{\mathbf{q}}{2}+\mathbf{k}%
,\frac{\omega}{2}+\omega_{k}\right] \nonumber\\
&  -g_{0}C_{+}\left[  \frac{\mathbf{q}}{2}-\mathbf{k},\frac{\omega}{2}%
-\omega_{k}\right]  G_{-}\left[  \frac{\mathbf{q}}{2}+\mathbf{k},\frac{\omega
}{2}+\omega_{k}\right]  \Bigg\}.
\end{align}
The key feature here is the mixing of the propagators for the charge and
density fields [dashed and solid lines in Eq.~\eqref{eqn:SE-diagrams}] which
prevents us from eliminating the linear drift terms $v_{\pm}k_{\parallel}$ in
both propagators simultaneously, unlike the case in the single-species DDS.
Substituting in the expressions in Eq.~(\ref{eq:correlations}), we evaluate
the integral over $\omega_{k}$ first. To simplify calculations, we dropped the
terms quartic in $\mathbf{q}$ (which would have provided a UV cutoff $\Lambda$
we introduce below). The result is%
\begin{align}
\Sigma_{-}&\approx\int 
\frac{\mathrm{d}\mathbf{k}}{(2\pi)^{2}}   \frac{\bar{N}\bar{\rho}g_{0}q_{\parallel}\left[g_{-}\left(  \frac{q_{\parallel}%
}{2}-k_{\parallel}\right)  -g_{0}\left(  \frac{q_{\parallel}}{2}+k_{\parallel
}\right) \right]  }{ 2\bar{D}^2(\frac{|\mathbf{q}|^{2}}{4}+|\mathbf{k}|^{2})-i\bar{D}\left[  \omega-\bar{v}q_{\parallel}+v_{d}k_{\parallel}\right]
}%
\end{align}
to leading order in the average density $\bar{\rho}$. After integrating
$k_{\bot}$ on the real line first, the remaining $\int k_{\parallel}$ is
na\"ively linearly divergent. However, by imposing the UV cutoff $\Lambda
=\pi/\ell$, we find the result in the main text:%
\begin{align}
\Sigma_{-}  &  \approx\frac{\bar{\rho}g_{0}\bar{N}v_{d}q_{\parallel}}%
{32\pi\bar{D}^{3}}\Bigg\{\left[  \frac{ q_{\parallel}%
}{ q_c}\,(g_{-}-g_{0})-i(g_{-}+g_{0})\right] \nonumber\\
&  \times\ln\left[  \frac{16\Lambda^{2}}{2i(\bar
{v}q_{\parallel}-\omega)/\bar{D}+q_c^2+  |\mathbf{q}|^{2} }\right]
+i(g_{-}+g_{0})\Bigg\}.
\end{align}
Here, $\bar{D}=(D_{+}+D_{-})/2$ is the average diffusivity,
$\bar{v}=(v_{+}+v_{-})/2$ the average velocity, $v_{d}=-\bar{\rho}\delta_{v}$
the velocity difference, $q_c=v_d/2\bar{D}$ the crossover wavenumber, and $\bar{N}=(N_{-}+N_{+})/2$ is the noise
correlation magnitude. The real part, $\Re\Sigma_{-}$, is given in Section V.
For completeness, we provide the imaginary part here:
\begin{align}
&  \Im\Sigma_{-}=\frac{\bar{\rho}g_{0}\bar{N}q_{\parallel}}{16\pi\bar{D}^{2}%
}\Bigg\{(g_{0}-g_{-})q_{\parallel}\tan^{-1}\left[  \frac{2(\bar
{v}q_{\parallel}-\omega)}{\bar{D}(q_c^2+|\mathbf{q}|^{2})}\right]
\nonumber\\
&  {}-\frac{(g_{0}+g_{-})q_c}{2}\ln\bigg[\frac{2^{12}\bar{D}^{2}\Lambda^{4}e^{-4}}{[ \bar{v}q_{\parallel}-\omega]^{2}+16\bar{D}^2[ q_c^2 + |\mathbf{q}|^{2}]^{2}}\bigg]\Bigg\}.
\end{align}

Meanwhile, the single integral associated with the noise correction for the charge field is 
\begin{align}
\eta_{-}  &  \approx\frac{\left[  \bar{N}\bar{\rho}g_{0}q_{\parallel}\right]
^{2}}{\bar{D}}\int\frac{\mathrm{d}\mathbf{k}}{(2\pi)^{2}}\,\nonumber\\
&  \frac{|\mathbf{q}|^{2}+4|\mathbf{k}|^{2}}{\left[  \omega-\bar
{v}q_{\parallel}-v_{d}k_{\parallel}\right]  ^{2}+\left[  \frac{\bar
{D}|\mathbf{q}|^{2}}{2}+2\bar{D}|\mathbf{k}|^{2}\right]  ^{2}} 
\end{align}
Similarly, after the $\omega_{k}$ integration, we find the self-energy for the
density field, represented by Eq.~\eqref{eq:Sigma+pix}, to be 
\begin{align}
\Sigma_{+}  &  \approx\frac{\bar{\rho}q_{\parallel}^{2}}{2}\int\frac
{\mathrm{d}\mathbf{k}}{(2\pi)^{2}}\Bigg\{\nonumber\\
&  \frac{N_{-}g_{0}g_{-}}{2D_{-}^{2}\left[
\frac{|\mathbf{q}|^{2}}{4}+|\mathbf{k}|^{2}\right] -iD_{-}(\omega-v_{-}q_{\parallel})  }\\
&  {}-\frac{N_{+}g_{+}^{2}}{ 2D_{+} 
^{2}\left[  \frac{|\mathbf{q}|^{2}}{4}+|\mathbf{k}|^{2}\right]-iD_{+}(\omega-v_{+}q_{\parallel})  } 
\Bigg\},\nonumber
\end{align}
leading to the result in Eq.~\eqref{eqn:Sigma+}. For completeness, we display
its real and imaginary parts:
\begin{align}
&  \Re\Sigma_{+}\approx\frac{\bar{\rho}q_{\parallel}^{2}}{32\pi}%
\Bigg\{  \frac{N_{-}g_{0}g_{-}}{D_{-}^{2}}\ln\left[  \frac{256\Lambda^{4}}{[\frac
{2}{D_{-}}(v_{-}q_{\parallel}-\omega)]^{2}+|\mathbf{q}|^{4}}\right]
\nonumber\\
&  {}-\frac{N_{+}g_{+}^{2}}{D_{+}^{2}}\ln\left[  \frac{256\Lambda^{4}}%
{[\frac{2}{D_{+}}(v_{+}q_{\parallel}-\omega)]^{2}+|\mathbf{q}|^{4}}\right]
\Bigg\}. \label{eq:ReSigmaP} 
\end{align}
and
\begin{align}
\Im\Sigma_{+}  &  \approx\frac{\bar{\rho}q_{\parallel}^{2}}{16\pi
}\Bigg\{   \frac{N_{+}g_{+}^{2}}{D_{+}^{2}}\tan^{-1}\left[  \frac{2(v_{+}q_{\parallel
}-\omega)}{D_{+}|\mathbf{q}|^{2}}\right] \nonumber\\
&  {}-\frac{N_{-}g_{0}g_{-}}{D_{-}^{2}}\tan^{-1}\left[  \frac{2(v_{-} 
q_{\parallel}-\omega)}{D_{-}|\mathbf{q}|^{2}}\right]  \Bigg\}.
\end{align}
Finally, the correction to the noise correlations, $\bar{N}_{+}=N_{+} 
\bar{\rho}|\mathbf{q}|^{2}+\eta_{+}$ [associated with Eq.~\eqref{eq:eta+pix}]
is given by
\begin{align}
\eta_{+} &
 \approx\frac{\bar{\rho}^{2}q_{\parallel}^{2}}{4}\int\frac{\mathrm{d} 
{\mathbf{k}}\,}{(2\pi)^{2}}\Bigg\{ \nonumber \\ &\sum_{\alpha=\pm}\frac{g_{\alpha} 
^{2}N_{\alpha}^{2}\left[  \frac{|\mathbf{q}|^{2}} 
{4}+|\mathbf{k}|^{2}\right]  }{\left(  \omega-v_{\alpha}k_{\parallel}\right)  ^{2}\frac{D_{\alpha}}{4}+D_{\alpha}^3\left[  \frac{|\mathbf{q}|^{2}}{4}
+|\mathbf{k}|^{2}\right]  ^{2}} \Bigg\},
\end{align}
which leads to the result in Eq.~\eqref{eq:eta+}.

\end{fmffile}

\bibliographystyle{apsrev}
  \bibliography{microe}

\end{document}